\documentclass[12pt,a4paper,twoside]{article} 

\usepackage{etoolbox} 
\usepackage[english]{babel}
\usepackage[T1]{fontenc}
\usepackage[utf8]{inputenc}
\usepackage{csquotes}
\usepackage{lmodern}
\usepackage[backend=biber,style=alphabetic,maxnames=10]{biblatex} 

\usepackage{nameref}
\usepackage{bookmark}
\usepackage{graphicx}
\usepackage[usenames,dvipsnames]{xcolor}
\usepackage{pgfplots} 
\usetikzlibrary{arrows,positioning,shapes,intersections,patterns,calc,fit,external} 
\usepgfplotslibrary{groupplots,fillbetween}
\usepackage{pgfplotstable}
\usepackage{subcaption}
\usepackage{caption}
\usepackage{tabularx}
\usepackage{booktabs}
\usepackage{multirow}
\usepackage{placeins}
\usepackage{amsmath}
\usepackage{amssymb}
\usepackage{amsthm}
\usepackage{mdframed}
\mdfsetup{nobreak=true}
\usepackage{mathtools}
\usepackage[eulergreek]{sansmath}
\usepackage{siunitx}
\usepackage{interval}
\intervalconfig{soft open fences,separator symbol =;}
\usepackage{balance}
\usepackage{color} 
\usepackage{colortbl} 
\usepackage{longtable}
\usepackage{enumerate}
\usepackage[nohyperlinks]{acronym} 
\usepackage{algpseudocode,algorithm}

\usepackage{hyperref}
\usepackage[noabbrev]{cleveref} 

\usepackage[left=2.5cm,right=2cm, vmargin=3cm]{geometry}
\usepackage{scrlayer-scrpage}
\usepackage{scrhack}

\newtheorem{assum}{Assumption}

\newmdtheoremenv[
  linecolor=black,
  roundcorner=5pt,
  linewidth=1pt,
  innertopmargin=2pt
]{thm}{Theorem}
\newmdtheoremenv[
  linecolor=black,
  roundcorner=5pt,
  linewidth=1pt,
  innertopmargin=2pt
]{prop}{Proposition}
\newmdtheoremenv[
  linecolor=black,
  roundcorner=5pt,
  linewidth=1pt,
  innertopmargin=2pt
]{lem}{Lemma}
\newmdtheoremenv[
  linecolor=black,
  roundcorner=5pt,
  linewidth=1pt,
  innertopmargin=2pt
]{cor}{Corollary}
\newtheorem{rem}{Remark}
\theoremstyle{definition}
\newtheorem{defn}{Definition}

\newmdtheoremenv[
nobreak=false,
  linecolor=black!5,
  backgroundcolor=black!5,
  roundcorner=5pt,
  linewidth=1pt,
  innertopmargin=2pt
]{exam}{Example}
\newcommand{\R}{\mathbb{R}}

\newcommand{\N}{\mathbb{N}}

\newcommand{\D}{\mathcal{D}}
\newcommand{\X}{\mathcal{X}}
\newcommand{\Z}{\mathcal{Z}}

\newcommand{\bm}[1]{{\boldsymbol{#1}}}

\newcommand{\Verts}[1]{{\left\Vert #1 \right\Vert}}

\DeclareMathOperator{\diag}{diag}
\DeclareMathOperator{\var}{var}
\DeclareMathOperator{\mean}{\mu}
\DeclareMathOperator{\Var}{\Sigma}
\DeclareMathOperator{\Mean}{\bm\mu}

\DeclareMathOperator{\prob}{p}
\DeclareMathOperator{\Prob}{P}
\newcommand{\ev}{\operatorname{E}}

\makeatletter
\DeclareRobustCommand{\rvdots}{%
  \vbox{
    \baselineskip4\p@\lineskiplimit\z@
    \kern-\p@
    \hbox{.}\hbox{.}\hbox{.}
  }}
  \makeatother
	\pgfplotsset{select coords between index/.style 2 args={
    x filter/.code={
        \ifnum\coordindex<#1\fi
        \ifnum\coordindex>#2\fi
    }
	}}

\newcommand{\GP}{\mathcal{GP}}

\newcommand{\h}{\bm{h}}

\newcommand{\x}{\bm x}

\newcommand{\xk}{\bm{x}_t}
\newcommand{\xkp}{\bm{x}_{t+1}}

\newcommand{\inputu}{\bm{u}}
\newcommand{\uk}{{\bm u}_t}
\newcommand{\yk}{\bm{y}_t}
\newcommand{\ykp}{\bm{y}_{t+1}}

\newcommand{\zk}{\bm{\zeta}_t}
\newcommand{\dyn}{\bm{f}}

\newcommand{\y}{\bm{y}}
\newcommand{\vxi}{\bm{\xi}}

\newcolumntype{M}{R@{${}={}$}L}
\crefname{propy}{Property}{Properties}
\crefname{exam}{Example}{Examples}
\crefname{chal}{Challenge}{Challenges}
\crefname{rem}{Remark}{Remarks}
\crefname{assum}{Assumption}{Assumptions}
\crefname{prop}{Proposition}{Propositions}
\crefname{cor}{Corollary}{Corollaries}
\crefname{lem}{Lemma}{Lemmas}
\crefname{thm}{Theorem}{Theorems}
\crefname{defn}{Definition}{Definitions}
\crefname{figure}{Fig.}{Fig.}
\Crefname{figure}{Figure}{Figures}
\crefname{table}{Table}{Tables}
\crefname{section}{Section}{Sections}
\crefname{chapter}{Chapter}{Chapters}
\crefname{equation}{}{}
\Crefname{equation}{Equation}{Equations}
\crefname{algorithm}{Algorithm}{Algorithms}

\pgfplotsset{compat=1.5}

\definecolor{TUMblue}{RGB}{0, 101, 189}
\definecolor{TUMwhite}{RGB}{255, 255, 255}
\definecolor{TUMblack}{RGB}{0, 0, 0}

\definecolor{TUMblue1}{RGB}{0, 82, 147}
\definecolor{TUMblue2}{RGB}{0, 51, 89}
\definecolor{TUMgray1}{RGB}{88, 88, 90}
\definecolor{TUMgray2}{RGB}{156, 157, 159}
\definecolor{TUMgray3}{RGB}{217, 218, 219}

\definecolor{TUMgray4}{RGB}{218, 215, 203}
\definecolor{TUMorange}{RGB}{227, 114, 34}
\definecolor{TUMgreen}{RGB}{162, 173, 0}
\definecolor{TUMlightblue1}{RGB}{152, 198, 234}
\definecolor{TUMlightblue2}{RGB}{100, 160, 200}

\definecolor{matlabcyanold}{RGB}{0, 255, 255}
\definecolor{matlabmagenta}{RGB}{255, 0, 255}

\definecolor{matlabblue}{RGB}{0, 113.985, 188.955}
\definecolor{matlaborange}{RGB}{216.75, 82.875, 24.99}
\definecolor{matlabyellow}{RGB}{236.895, 176.97, 31.875}
\definecolor{matlabpurple}{RGB}{125.97, 46.92, 141.78}
\definecolor{matlabgreen}{RGB}{118.83, 171.87, 47.94}
\definecolor{matlabcyan}{RGB}{76.755, 189.975, 237.915}
\definecolor{matlabred}{RGB}{161.925, 19.89, 46.92}
\graphicspath{{fig/}{logos/}}
\begin{document}
\thispagestyle{empty}
\enlargethispage{4.5cm} 
\begin{center}
\phantom{u}
\vspace{0.5cm}
\Huge{\sc An Introduction to\\Gaussian Process Models}\\
\vspace{1.5cm}
                                 \large{
                                           

						\vspace{2cm}
					by\\
					Thomas Beckers\\
					t.beckers@tum.de\\
					\vspace{1cm}	
					Abstract}
\end{center}
Within the past two decades, Gaussian process regression has been increasingly used for modeling dynamical systems due to some beneficial properties such as the bias variance trade-off and the strong connection to Bayesian mathematics. As data-driven method, a Gaussian process is a powerful tool for nonlinear function regression without the need of much prior knowledge. In contrast to most of the other techniques, Gaussian Process modeling provides not only a mean prediction but also a measure for the model fidelity. In this article, we give an introduction to Gaussian processes and its usage in regression tasks of dynamical systems. Try it yourself: \href{https://gpr.tbeckers.com}{gpr.tbeckers.com}
\begin{center}
					\large{\vspace{1cm}	
					Original Work: April, 2020\\
					Current Revision: Feb 10, 2021\\
					\vspace{3cm}
					Chair of\\
					Information-oriented Control\\
					Technical University of Munich}
\end{center}

\newpage
\thispagestyle{empty}
\tableofcontents 
\newpage
\section{Introduction}
A Gaussian process (GP) is a stochastic process that is in general a collection of random variables indexed by time or space. Its special property is that any finite collection of these variables follows a multivariate Gaussian distribution. Thus, the GP is a distribution over infinitely many variables and, therefore, a distribution over functions with a continuous domain. Consequently, it describes a probability distribution over an infinite dimensional vector space. For engineering applications, the GP has gained increasing attention as supervised machine learning technique, where it is used as prior probability distribution over functions in Bayesian inference. The inference of continuous variables leads to Gaussian process regression (GPR) where the prior GP model is updated with training data to obtain a posterior GP distribution. Historically, GPR was used for the prediction of time series, at first presented by Wiener and Kolmogorov in the 1940's. Afterwards, it became increasingly popular in geostatistics in the 1970's, where GPR is known as \emph{kriging}. Recently, it came back in the area of machine learning~\cite{neal1996bayesian,williams1996gaussian}, especially boosted by the rapidly increasing computational power.\\
In this article, we present background information about GPs and GPR, mainly based on~\cite{rasmussen2006gaussian}, focusing on the application in control. We start with an introduction of GPs, explain the role of the underlying kernel function and show its relation to reproducing kernel Hilbert spaces. Afterwards, the embedding in dynamical systems and the interpretation of the model uncertainty as error bounds is presented. Several examples are included for an intuitive understanding in addition to the formal notation.
\section{Gaussian Processes}
Let~$(\Omega_\text{ss}, \mathcal{F}_\sigma,P)$ be a probability space with the sample space~$\Omega_\text{ss}$, the corresponding~$\sigma$-algebra~$\mathcal{F}_\sigma$ and the probability measure~$\text{P}$. The index set is given by~$\Z \subseteq \R^{n_z}$ with positive integer~${n_z}$. Then, a function~$f_{\text{GP}}(\bm{z}, \omega_\text{ss})$, which is a measurable function of~$\omega_\text{ss}\in\Omega_\text{ss}$ with index~$\bm{z}\in\Z$, is called a stochastic process. The function~$f_{\text{GP}}(\bm{z}, \omega_\text{ss})$ is a random variable on~$\Omega_\text{ss}$ if~$\bm{z}\in\Z$ is specified. It is simplified written as~$f_{\text{GP}}(\bm{z})$. A GP is a stochastic process which is fully described by a mean function~$m\colon\Z\to\R$ and covariance function~$k\colon \Z\times \Z\to\R$ such that
\begin{align}
	f_{\text{GP}}(\bm{z}) &\sim \GP\left(m(\bm{z}),k(\bm{z},\bm{z}^\prime)\right)\\
	\begin{split}
	m(\bm{z})&=\ev\left[f_{\text{GP}}(\bm{z})\right]\\
	k(\bm{z},\bm{z}^\prime)&=\ev\left[\left(f_{\text{GP}}(\bm{z})-m(\bm{z})\right)\left(f_{\text{GP}}(\bm{z}^\prime)-m(\bm{z}^\prime)\right)\right]
	\end{split}
\end{align}
with~$\bm{z},\bm{z}^\prime\in\Z$. The covariance function is a measure for the correlation of two states~$(\bm{z},\bm{z}^\prime)$ and is called \textit{kernel} in combination with GPs. Even though no analytic description of the probability density function of the GP exists in general, the interesting property is that any finite collection of its random variables~$\{f_{\text{GP}}(\bm{z}_1),\ldots,f_{\text{GP}}(\bm{z}_{n_{\text{GP}}})\}$ follows a~$n_{\text{GP}}$-dimensional multivariate Gaussian distribution. As a GP defines a distribution over functions, each realization is also a function over the index set~$\Z$.
\begin{exam}
A GP~$f_{\text{GP}}(t_c)\sim \GP\left(m(t_c),k(t_c,t_c^\prime)\right)$ with time~$t_c\in\R_{\geq 0}$, where
\begin{align}
m(t_c)=1\si{\ampere},\quad k(t_c,t_c^\prime)=\begin{cases}(0.1\si{\ampere})^{2}&t_c=t_c^\prime\\(0\si{\ampere})^{2}&t_c\neq t_c^\prime\end{cases}\notag
\end{align}
describes a time-dependent electric current signal with Gaussian white noise with a standard deviation of~$\SI{0.1}{\ampere}$ and a mean of~$\SI{1}{\ampere}$.
\end{exam}
\subsection{Gaussian Process Regression}
\label{sec2:sec:GPR}
The GP can be utilized as prior probability distribution in Bayesian inference, which allows to perform function regression. Following the Bayesian methodology, new information is combined with existing information: using Bayes’ theorem, the prior is combined with new data to obtain a posterior distribution. The new information is expressed as training data set~$\D=\{X,Y\}$. It contains the input values~$X=[\x_\text{dat}^{\{1\}},\x_\text{dat}^{\{2\}},\ldots,\x_\text{dat}^{\{n_\D\}}]\in\Z^{1\times {n_\D}}$ and output values~$Y=[\tilde{y}_\text{dat}^{\{1\}},\tilde{y}_\text{dat}^{\{2\}},\ldots,\tilde{y}_\text{dat}^{\{{n_\D}\}}]^\top\in\R^{{n_\D}}$, where
\begin{align}
\tilde{y}_\text{dat}^{\{i\}}=f_{\text{GP}}(\x_\text{dat}^{\{i\}})+\nu
\end{align}
for all~$i=1,\ldots,n_\D$. 
The output data might be corrupted by Gaussian noise~$\nu\sim\mathcal{N}(0,\sigma_n^2)$. 
\begin{rem}
Note that we always use the standard notation~$X$ for the input training data and~$Y$ for the output training data throughout this report. 
\end{rem}
As any finite subset of a GP follows a multivariate Gaussian distribution, we can write the joint distribution
\begin{align}
\begin{bmatrix} Y\vphantom{\begin{bmatrix}m(\x_\text{dat}^{\{1\}})\\\vdots\\ m(\x_\text{dat}^{\{{n_\D}\}})\end{bmatrix}} \\ f_\text{GP}(\bm{z}^*) \end{bmatrix}\sim \mathcal{N} \left(\begin{bmatrix}m(\x_\text{dat}^{\{1\}})\\\vdots\\ m(\x_\text{dat}^{\{{n_\D}\}})\\m(\bm{z}^*)\end{bmatrix}, \begin{bmatrix} K(X,X)+\sigma_n^2 I_{n_\D}\vphantom{\begin{bmatrix}m(\x_\text{dat}^{\{1\}})\\\vdots\\ m(\x_\text{dat}^{\{{n_\D}\}})\end{bmatrix}} & \bm k(\bm{z}^*,X)\\ \bm k(\bm{z}^*,X)^\top & k(\bm{z}^*,\bm{z}^*) \end{bmatrix}\right)\label{sec2:for:joint_dist}
\end{align} 
for any arbitrary test point~$\bm{z}^*\in\Z$. The function~$m\colon\Z\to\R$ denotes the mean function. The matrix function~$K\colon\Z^{1\times {n_\D}}\times \Z^{1\times {n_\D}}\to\R^{{n_\D}\times {n_\D}}$ is called the covariance or \textit{Gram matrix} with
\begin{align}
K_{j,l}(X,X)= k(X_{:, l},X_{:, j})\text{ for all }j,l\in\lbrace 1,\ldots,{n_\D}\rbrace\label{sec2:for:gp_K}
\end{align}
where each element of the matrix represents the covariance between two elements of the training data~$X$. The expression~$X_{:, l}$ denotes the~$l$-th column of~$X$. For notational simplification, we shorten~$K(X,X)$ to~$K$ when necessary. The vector-valued kernel function~$\bm k\colon\Z\times \Z^{1\times {n_\D}}\to\R^{n_\D}$ calculates the covariance between the test input~$\bm{z}^*$ and the input training data~$X$, i.e.,
\begin{align}
\bm k(\bm{z}^*,X) = [k(\bm{z}^*,X_{:, 1}),\ldots,k(\bm{z}^*,X_{:, {n_\D}})]^\top . 
\end{align}
To obtain the posterior predictive distribution of~$f_\text{GP}(\bm{z}^*)$, we condition on the test point~$\bm{z}^*$ and the training data set~$\D$ given by
\begin{align}
\prob(f_\text{GP}(\bm{z}^*)\vert\bm{z}^*,\D)=\frac{\prob(f_\text{GP}(\bm{z}^*),Y\vert X,\bm{z}^*)}{\prob(Y\vert X)}.
\end{align}
Thus, the conditional posterior Gaussian distribution is defined by the mean and the variance
\begin{align}
	\mean(f_\text{GP}(\bm{z}^*)\vert \bm{z}^*,\mathcal D)&=m(\bm{z}^*)+\bm k(\bm{z}^*,X)^\top (K+\sigma_n^2 I_{n_\D})^{-1}\left(Y-[m(X_{:,1}),\ldots,m(X_{:,{n_\D}})]^\top\right)\notag\\
	\var(f_\text{GP}(\bm{z}^*)\vert \bm{z}^*,\mathcal D)&=k(\bm{z}^*,\bm{z}^*)-\bm k(\bm{z}^*,X)^\top (K+\sigma_n^2 I_{n_\D})^{-1} \bm k(\bm{z}^*,X).\label{sec2:for:gp_post}
\end{align}
A detailed derivation of the posterior mean and variance based on the joint distribution~\cref{sec2:for:joint_dist} can be found in~\cref{app:1}. Analyzing~\cref{sec2:for:gp_post} we can make the following observations:\\
\textbf{i)} The mean prediction can be written as
\begin{align}
\mean(f_\text{GP}(\bm{z}^*)\vert \bm{z}^*,\mathcal D)=m(\bm{z}^*)+\sum_{j=1}^{n_\D} \alpha_j k(\bm{z}^*,X_{:,j})
\end{align}
with~$\bm \alpha=(K+\sigma_n^2 I_{n_\D})^{-1}\left(Y-[m(X_{:,1}),\ldots,m(X_{:,{n_\D}})]^\top\right)\in\R^{n_\D}$. That formulation highlights the data-driven characteristic of the GPR as the posterior mean is a sum of kernel functions and its number grows with the number~${n_\D}$ of training data.\\
\textbf{ii)} The variance does not depend on the observed data, but only on the inputs, which is a property of the Gaussian distribution. The variance is the difference between two terms: The first term~$k(\bm{z}^*,\bm{z}^*)$ is simply the prior covariance from which a (positive) term is subtracted, representing the information the observations contain about the function. The uncertainty of the prediction, expressed in the variance, holds only for~$f_\text{GP}(\bm{z}^*)$ and does not consider the noise in the training data. For this purpose, an additional noise term~$\sigma_n^2 I_{{n_\D}}$ can be added to the variance in \cref{sec2:for:gp_post}. Finally, \cref{sec2:for:gp_post} clearly shows the strong dependence of the posterior mean and variance on the kernel~$k$ that we will discuss in depth in~\cref{sec2:sec:kernel}.
\begin{exam}
We assume a GP with zero mean and a kernel function given by
\begin{align}
k(z,z^\prime)=0.3679^2\exp\left(-\frac{(z-z^\prime)^2}{2\cdot 2.7183^2}\right)\notag
\end{align}
as prior distribution. The training data set~$\D$ is assumed to be
\begin{align}
X=\begin{bmatrix}
1 & 3 & 6 &10
\end{bmatrix},\quad
Y=\begin{bmatrix}
0 & -0.3 & 0.3 & -0.2
\end{bmatrix}^\top,\notag
\end{align}
where the output is corrupted by Gaussian noise with~$\sigma_n=0.0498$ standard deviation and the test point is assumed to be~$z^*=5$. According to~\crefrange{sec2:for:gp_K}{sec2:for:gp_post} the Gram matrix~$K(X,X)$ is calculated as
\begin{align}
K(X,X)=\begin{bmatrix}
	0.1378 &   0.1032  &  0.0249  &  0.0006\\
    0.1032 &   0.1378  &  0.0736  &  0.0049\\
    0.0249 &   0.0736  &  0.1378  &  0.0458\\
    0.0006 &   0.0049  &  0.0458  &  0.1378
\end{bmatrix}\notag
\end{align}
and the kernel vector~$\bm k(z^*,X)$ and~$k(z^*,z^*)$ are obtained to be
\begin{align}
\bm k(z^*,X)&=\begin{bmatrix}
    0.0458  &  0.1032  &  0.1265  &  0.0249
\end{bmatrix}\notag\\
k(z^*,z^*)&=0.1378.\notag
\end{align}
Finally, with~\cref{sec2:for:gp_post}, we compute the predicted mean and variance for~$f_\text{GP}(z^*)$
\begin{align}
	\mean(f_\text{GP}(z^*)\vert z^*,\D)=0.0278,\quad \var(f_\text{GP}(z^*)\vert z^*,\D)=0.0015,\notag
\end{align}
which is equivalent to a~$2\sigma$-standard deviation of~$0.0775$. \Cref{sec2:fig:prior_posterior} shows the prior distribution (left), the posterior distribution with two training points (black crosses) in the middle, and the posterior distribution given the full training set~$\D$ (right). The solid red line is the mean function and the gray shaded area indicates the~$2\sigma$-standard deviation. Five realizations (dashed lines) visualize the character of the distribution over functions.
	\begin{center}
		\tikzsetnextfilename{section2_prior_posterior}
		\captionsetup{type=figure}\begin{tikzpicture}
	\begin{axis}[
	  name=plot1,
	  axis lines=left,
	  ylabel={Output space},
	  font={\sffamily},
	  line width=1pt,
	  grid = none,
	  height=5cm,
	  width=6.1cm,
	  xmin=0, xmax=11,ymin=-0.7,ymax=0.9,
	 ylabel style={at={(-0.2,0.5)}}]
			 \addplot[name path=varp1, color=gray,opacity=0.3, no marks] table [x index=0,y expr=\thisrowno{1}+2*(\thisrowno{2})]{data/section2/figure3_var.dat};
			 \addplot[name path=varm1, color=gray,opacity=0.3, no marks] table [x index=0,y expr=\thisrowno{1}-2*(\thisrowno{2})]{data/section2/figure3_var.dat};
			 \addplot[gray,opacity=0.5] fill between[ of = varm1 and varp1]; 
			 \addplot[mark=+,color=black, only marks,mark size=6,line width=1pt,select coords between index={0}{0}] table [x index=0,y index=1]{data/section2/figure3_data.dat};
  			 \addplot[color=green,dashed,line width=1pt] table [x index=0,y index=1]{data/section2/figure3_samples.dat};
			 \addplot[color=purple,dashed,line width=1pt] table [x index=0,y index=2]{data/section2/figure3_samples.dat};
			 \addplot[color=orange,dashed,line width=1pt] table [x index=0,y index=3]{data/section2/figure3_samples.dat};
			 \addplot[color=blue,dashed,line width=1pt] table [x index=0,y index=4]{data/section2/figure3_samples.dat};
			 \addplot[color=yellow,dashed,line width=1pt] table [x index=0,y index=5]{data/section2/figure3_samples.dat};
  			 \addplot[color=red,line width=1pt] table [x index=0,y index=1]{data/section2/figure3_var.dat};	 
	\end{axis}
	\begin{axis}[
	  name=plot2,
	  xshift=-1mm,
	  at=(plot1.right of south east), anchor=left of south west,
	  axis lines=left,
	  xlabel={Input space},
	  font={\sffamily},
	  line width=1pt,
	  grid = none,
	  height=5cm,
	  width=6.1cm,
	  yticklabels={,,},
	  xmin=0, xmax=11,ymin=-0.7,ymax=0.9]
			 \addplot[name path=varp1, color=gray,opacity=0.3, no marks] table [x index=0,y expr=\thisrowno{5}+2*(\thisrowno{6})]{data/section2/figure3_var.dat};
			 \addplot[name path=varm1, color=gray,opacity=0.3, no marks] table [x index=0,y expr=\thisrowno{5}-2*(\thisrowno{6})]{data/section2/figure3_var.dat};
			 \addplot[gray,opacity=0.5] fill between[ of = varm1 and varp1]; 
			 \addplot[mark=+,color=black, only marks,mark size=6,line width=1pt,select coords between index={0}{2}] table [x index=0,y index=1]{data/section2/figure3_data.dat};
  			 \addplot[color=green,dashed,line width=1pt] table [x index=0,y index=11]{data/section2/figure3_samples.dat};
			 \addplot[color=purple,dashed,line width=1pt] table [x index=0,y index=12]{data/section2/figure3_samples.dat};
			 \addplot[color=orange,dashed,line width=1pt] table [x index=0,y index=13]{data/section2/figure3_samples.dat};
			 \addplot[color=blue,dashed,line width=1pt] table [x index=0,y index=14]{data/section2/figure3_samples.dat};
			 \addplot[color=yellow,dashed,line width=1pt] table [x index=0,y index=15]{data/section2/figure3_samples.dat};
  			 \addplot[color=red,line width=1pt] table [x index=0,y index=5]{data/section2/figure3_var.dat};		 
	\end{axis}
		\begin{axis}[
	  name=plot3,
	  at=(plot2.right of south east), anchor=left of south west,
	  xshift=-1mm,
	  axis lines=left,
	  line width=1pt,
	  grid = none,
	  height=5cm,
	  width=6.1cm,
	  xticklabels={,$0$,$z^*\!=\! 5$,$10$},
	  yticklabels={,,},
	  xmin=0, xmax=11,ymin=-0.7,ymax=0.9]
			 \addplot[name path=varp1, color=gray,opacity=0.3, no marks] table [x index=0,y expr=\thisrowno{9}+2*(\thisrowno{10})]{data/section2/figure3_var.dat};
			 \addplot[name path=varm1, color=gray,opacity=0.3, no marks] table [x index=0,y expr=\thisrowno{9}-2*(\thisrowno{10})]{data/section2/figure3_var.dat};
			 \addplot[gray,opacity=0.5] fill between[ of = varm1 and varp1]; 
			 \addplot[mark=+,color=black, only marks,mark size=6,line width=1pt,select coords between index={0}{4}] table [x index=0,y index=1]{data/section2/figure3_data.dat};
  			 \addplot[color=green,dashed,line width=1pt] table [x index=0,y index=21]{data/section2/figure3_samples.dat};
			 \addplot[color=purple,dashed,line width=1pt] table [x index=0,y index=22]{data/section2/figure3_samples.dat};
			 \addplot[color=orange,dashed,line width=1pt] table [x index=0,y index=23]{data/section2/figure3_samples.dat};
			 \addplot[color=blue,dashed,line width=1pt] table [x index=0,y index=24]{data/section2/figure3_samples.dat};
			 \addplot[color=yellow,dashed,line width=1pt] table [x index=0,y index=25]{data/section2/figure3_samples.dat};
  			 \addplot[color=red,line width=1pt] table [x index=0,y index=9]{data/section2/figure3_var.dat};	 
	\end{axis}
	\end{tikzpicture} 
		\vspace{-0.5cm}
		
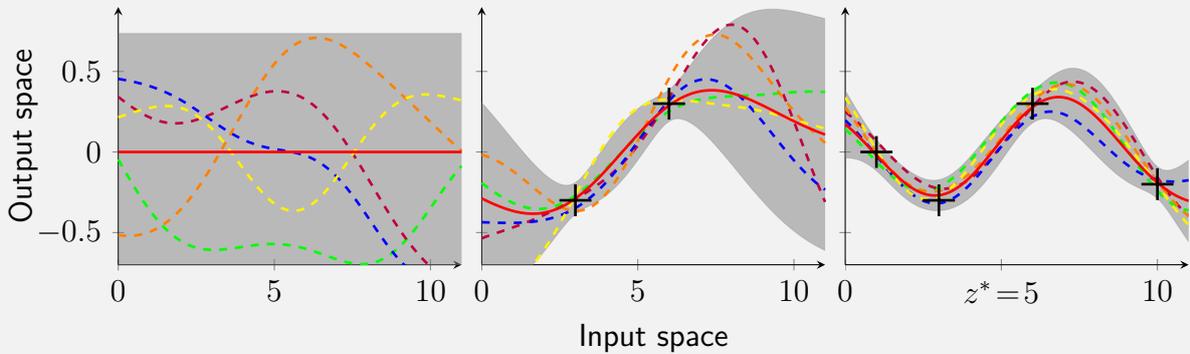
\captionof{figure}{The prior distribution of a GP is updated with data that leads to the posterior distribution.}
		\label{sec2:fig:prior_posterior}
	\end{center}
\end{exam}
\subsection{Multi-output Regression}
So far, the GP regression allows functions with scalar outputs as in~\cref{sec2:for:gp_post}. For the extension to vector-valued outputs, multiple approaches exist: i) Extending the kernel to multivariate outputs~\cite{MAL-036}, ii) adding the output dimension as training data~\cite{berkenkamp2017safe} or iii) using separated GPR for each output~\cite{rasmussen2006gaussian}. While the first two approaches set a prior on the correlation between the output dimensions, the latter disregards a correlation without loss of generality. Following the approach in iii), the previous definition of the training set~$\D$ is extended to a vector-valued output with
\begin{align}
X=[\x_\text{dat}^{\{1\}},\x_\text{dat}^{\{2\}},\ldots,\x_\text{dat}^{\{n_\D\}}]\in\Z^{1\times {n_\D}},\quad Y=[\tilde{\y}_\text{dat}^{\{1\}},\tilde{\y}_\text{dat}^{\{2\}},\ldots,\tilde{\y}_\text{dat}^{\{{n_\D}\}}]^\top\in\R^{{n_\D}\times {n_{y\text{dat}}}},
\end{align}
where~${n_{y\text{dat}}}\in\N$ is the dimension of the output and the vector-valued GP is defined by
\begin{align}
\bm{f}_{\text{GP}}(\bm{z}) &\sim \begin{cases}\GP\big(m^1(\bm{z}),k^1(\bm{z},\bm{z}^\prime)\big)\\ \hphantom{aaaaa}\vdots\hphantom{aaaaa}\vdots \\ \GP\big(m^{{n_{y\text{dat}}}}(\bm{z}),k^{{n_{y\text{dat}}}}(\bm{z},\bm{z}^\prime)\big)\end{cases}\\
\bm{m}(\bm{z})&\coloneqq\left[m^1(\bm{z}),\ldots,m^{{n_{y\text{dat}}}}(\bm{z})\right]^\top\label{sec2:for:multigpdef}
\end{align}
Following~\crefrange{sec2:for:joint_dist}{sec2:for:gp_post}, we obtain for the predicted mean and variance
\begin{align}
	\mean(f_{\text{GP},i}(\bm{z}^*)\vert \bm{z}^*\!,\mathcal D)&=m^i(\bm{z}^*)+\bm k^i(\bm{z}^*,X)^\top (K^i\!+\!\sigma_{n,i}^2 I_{n_\D})^{-1}\left(Y_{:,i}\!-\![m^i(X_{:,1}),\ldots,m^i(X_{:,{n_\D}})]^\top\right)\notag\\
	\var(f_{\text{GP},i}(\bm{z}^*)\vert \bm{z}^*\!,\mathcal D)&=k^i(\bm{z}^*,\bm{z}^*)-\bm k^i(\bm{z}^*,X)^\top (K^i\!+\!\sigma_{n,i}^2 I_{n_\D})^{-1} \bm k^i(\bm{z}^*,X)\label{sec2:for:gp_post_vec}
\end{align}
for each output dimension~$i\in\{1,\ldots,{n_{y\text{dat}}}\}$ with respect to the kernels~$k^1,\ldots,k^{{n_{y\text{dat}}}}$. The variable~$\sigma_{n,i}$ denotes the standard deviation of the Gaussian noise that corrupts the~$i$-th dimension of the output measurements. The~${n_{y\text{dat}}}$ components of~$\bm{f}_{\text{GP}}\vert \bm{z}^*,\mathcal D$ are combined into a multi-variable Gaussian distribution with
\begin{align}
\begin{split}
	\Mean(\bm{f}_{\text{GP}}\vert \bm{z}^*,\mathcal D)&=[\mean(f_{\text{GP},1}\vert \bm{z}^*,\mathcal D),\ldots,\mean(f_{\text{GP},{n_{y\text{dat}}}}\vert \bm{z}^*,\mathcal D)]^\top\\
	\Var(\bm{f}_{\text{GP}}\vert \bm{z}^*,\mathcal D)&=\diag\left(\var(f_{\text{GP},1}\vert \bm{z}^*,\mathcal D),\ldots,\var(f_{\text{GP},{n_{y\text{dat}}}}\vert \bm{z}^*,\mathcal D)\right),\label{for:multigp}
	\end{split}
\end{align}
where~$\Var(\bm{f}_{\text{GP}}\vert \bm{z}^*,\mathcal D)$ denotes the posterior variance matrix. This formulation allows to use a GP prior on vector-valued functions to perform predictions for test points~$\bm{z}^*$. This approach treats each output dimension separately which is mostly sufficient and easy-to-handle. An alternative approach is to include the dimension as additional input, e.g., as in~\cite{berkenkamp2017safe}, with the benefit of a single GP at the price of loss of interpretability. For highly correlated output data, a multi-output kernel might be beneficial, see~\cite{MAL-036}.
\begin{rem}
Without specific knowledge about a trend in the data, the prior mean functions~$m^1,\ldots,m^{{n_{y\text{dat}}}}$ are often set to zero, see~\cite{rasmussen2006gaussian}. Therefore, we set the mean functions to zero for the remainder of the report if not stated otherwise.
\end{rem}
\subsection{Kernel-based View}
\label{sec2:sec:kbv}
\begin{figure}[t]
\begin{center}
\tikzsetnextfilename{section2_kernel_trick}
	\begin{tikzpicture}[scale=1,>=latex]
	\def\so{4}
	\def\rwi{2}
	\def\rhe{1.3}
	\def\rx{1.5}
	\def\ry{1.6}
	\def\centerarc[#1](#2)(#3:#4:#5)
    { \draw[#1] ($(#2)+({#5*cos(#3)},{#5*sin(#3)})$) arc (#3:#4:#5); }
	\coordinate (y) at (0,2);
    \coordinate (x) at (2,0);
    \draw[<->] (y) node[above] {} -- (0,0) --  (x) node[right] {};
    \foreach \i in {0,...,3}
{
    \draw ({0.4*sin(\i/2*pi r)+1.2},{0.4*cos(\i/2*pi r)+1.2}) circle (2pt); 
}
    \foreach \i in {1,...,10}
{
	\node [draw,rectangle,minimum width=2pt,minimum height=2pt,inner sep=0pt,fill](c\i) at ({0.8*sin(\i/5*pi r)+1.2},{0.8*cos(\i/5*pi r)+1.2}){};
}
	\coordinate (y1) at (\so,2);
    \coordinate (x1) at (2+\so,0);
    \coordinate (z1) at (1.5+\so,1.5);
    \draw[<->] (y1) node[] {} -- (\so,0) --  (x1) node[] {};
    \draw[->]  (\so,0) --  (z1) node[] {};
    \draw[fill=white] (\so+1.1,1.1) circle (1cm); 
    \foreach \i in {27,...,31}
	{
		\node [draw,rectangle,minimum width=2pt,minimum height=2pt,inner sep=0pt,fill](c\i) at ({2.7*sin(\i/23*pi r)+3+\so},{2.7*cos(\i/23*pi r)+3}){};
	}
	\draw (\so+1.74,1.7) circle (2pt); 
	\draw (\so+1.54,1.85) circle (2pt); 
	\centerarc[red,thick](3+\so,3)(204:246:2.3)
	\centerarc[red,thick,dashed](\so,0)(24:66:2.3)
	\draw [->, thick] (2.5,1) to [out=40,in=150] node[auto] {$\phi$} (\so-0.5,1);
\end{tikzpicture}
	\caption{The mapping~$\bm{\phi}$ transforms the data points into a feature space where linear regressors can be applied to predict the output.}
	\label{fig:kernel_trick}
	\vspace{-0.5cm}
\end{center}
\end{figure}
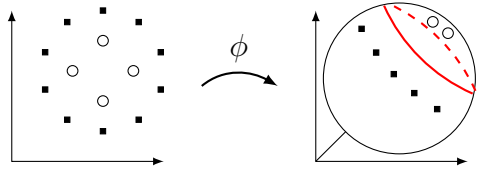

In~\cref{sec2:sec:GPR}, we target the GPR from a Bayesian perspective. However, for some applications of GPR a different point of view is beneficial; namely from the kernel perspective. In the following, we derive GPR from linear regression that is extended with a kernel transformation. In general, the prediction of parametric models is based on a parameter vector~$\bm{w}$ which is typically learned using a set of training data points. In contrast, non-parametric models typically maintain at least a subset of the training data points in memory in order to make predictions for new data points. Many linear models can be transformed into a dual representation where the prediction is based on a linear combination of kernel functions. The idea is to transform the data points of a model to an often high-dimensional feature space where a linear regression can be applied to predict the model output, as depicted in~\cref{fig:kernel_trick}. For a nonlinear feature map~$\bm{\phi}\colon\Z\to\mathcal{F}$, where~$\mathcal{F}$ is a~$n_\phi\in\N\cup\{\infty\}$ dimensional Hilbert space, the kernel function is given by the inner product~$k(\bm{z},\bm{z}^\prime)=\langle \bm{\phi}(\bm{z}),\bm{\phi}(\bm{z}^\prime)\rangle,\forall\bm{z},\bm{z}^\prime\in\Z$.
Thus, the kernel implicitly encodes the way the data points are transformed into a higher dimensional space. The formulation as inner product in a feature space allows to extend many standard regression methods. Also the GPR can be derived using a standard linear regression model 
\begin{align}
f_\text{lin}(\bm{z})=\bm{z}^\top\bm{w},\quad \tilde{y}_\text{dat}^{\{i\}}=f_{\text{GP}}(\x_\text{dat}^{\{i\}})+\nu
\end{align}
where~$\bm{z}\in\Z$ is the input vector,~$\bm{w}\in\R^{n_z}$ the vector of weights with~$n_z=\dim(\Z)$ and~$f_\text{lin}\colon\Z\to\R$ the unknown function. The observed value~$\tilde{y}_\text{dat}^{\{i\}}\in\R$ for the input~$\x_\text{dat}^{\{i\}}\in\Z$ is corrupted by Gaussian noise~$\nu\sim\mathcal{N}(0,\sigma_n^2)$ for all~$i=1,\ldots,n_\D$. The analysis of this model is analogous to the standard linear regression, i.e., we put a prior on the weights such that~$\bm{w}\sim\mathcal{N}(\bm{0},\Sigma_p)$ with~$\Sigma_p\in\R^{n_z\times n_z}$. Based on~${n_\D}$ collected training data points as defined in~\cref{sec2:sec:GPR}, that leads to the well known linear Bayesian regression
\begin{align}
\prob(f_\text{lin}(\bm{z}^*)\vert\bm{z}^*, \D)=\mathcal{N}\big(\underbrace{\frac{1}{\sigma_n^2}{\bm{z}^*}^\top A_\text{lin}^{-1} X Y}_{\mean(f_\text{lin}(\bm{z}^*)\vert\bm{z}^*, \D)},\underbrace{\vphantom{\frac{1}{\sigma_n^2}}{\bm{z}^*}^\top A_\text{lin}^{-1}{\bm{z}^*}}_{\var(f_\text{lin}(\bm{z}^*)\vert\bm{z}^*, \D)}\big)\label{sec2:for:lin}
\end{align}
where~$A_\text{lin}=\sigma_n^{-2}XX^\top+\Sigma_p^{-1}$. Now, using the feature map~$\bm{\phi}(\bm{z})$ instead of~$\bm{z}$ directly, leads to~$f_\text{GP}(\bm{z})=\bm{\phi}(\bm{z})^\top\check{\bm{w}}$ with~$\check{\bm{w}}\sim\mathcal{N}(\bm{0},\check{\Sigma}_p),\check{\Sigma}_p\in\R^{n_\phi\times n_\phi}$. As long as the projections are fixed functions, i.e., independent of the parameters~$w$, the model is still linear in the parameters and, thus, analytically tractable. In particular, the Bayesian regression~\cref{sec2:for:lin} with the mapping~$\bm{\phi}(\bm{z})$ can be written as
\begin{align}
(f_\text{GP}(\bm{z}^*)\vert\bm{z}^*,\D)\sim\mathcal{N}\left(\frac{1}{\sigma_n^2}\bm{\phi}(\bm{z}^*)^\top A_\text{GP}^{-1} \left[\bm{\phi}(X_{:,1});\ldots;\bm{\phi}(X_{:,{n_\D}})\right] Y,\bm{\phi}(\bm{z}^*)^\top A_\text{GP}^{-1}\bm{\phi}(\bm{z}^*)\right),
\end{align}
with the matrix $A_\text{GP}\in\R^{n_\phi\times n_\phi}$ given by
\begin{align}
A_\text{GP}=\sigma_n^{-2}\left[\bm{\phi}(X_{:,1});\ldots;\bm{\phi}(X_{:,{n_\D}})\right]\left[\bm{\phi}(X_{:,1});\ldots;\bm{\phi}(X_{:,{n_\D}})\right]^\top+\check{\Sigma}_p^{-1}.
\end{align}
This equation can be simplified and rewritten to
\begin{align}
(f_\text{GP}(\bm{z}^*)\vert\bm{z}^*,\D)\sim\mathcal{N}\big(\underbrace{\bm{k}(\bm{z}^*,X)^\top K^{-1}Y}_{\mean(f_\text{GP}(\bm{z}^*)\vert\bm{z}^*,\D)},\underbrace{k(\bm{z}^*,\bm{z}^*)-\bm{k}(\bm{z}^*,X)^\top K^{-1}\bm{k}(\bm{z}^*,X)}_{\var(f_\text{GP}(\bm{z}^*)\vert\bm{z}^*,\D)}\big),\label{sec2:for:gp2}
\end{align}
with~$k(\bm{z},\bm{z}^\prime)=\bm\phi(\bm{z})^\top\check{\Sigma}_p\bm\phi(\bm{z}^\prime)$ that equals~\cref{sec2:for:gp_post}. The fact that in~\cref{sec2:for:gp2} the feature map~$\bm{\phi}(\bm{z})$ is not needed is known as the \emph{kernel trick}. This trick is also used in other kernel-based models, e.g., support vector machines (SVM), see \cite{steinwart2008support} for more details.
\subsection{Reproducing Kernel Hilbert Space}
\label{sec2:sec:RKHS}
Even though a kernel neither uniquely defines the feature map nor the feature space, one can always construct a canonical feature space, namely the \textit{reproducing kernel Hilbert space} (RKHS) given a certain kernel. After the introduction of the theory, illustrative examples for an intuitive understanding are presented. We will now formally present this construction procedure, starting with the concept of Hilbert spaces, following~\cite{bhujwalla2016impact}: A Hilbert space~$\mathcal{F}$ represents all possible realizations of some class of functions, for example all functions of continuity degree~$i$, denoted by~$\mathcal{C}^i$. Moreover, a Hilbert space is a vector space such that any function~$f_\mathcal{F}\in\mathcal{F}$ must have a non-negative norm,~$\Vert f_\mathcal{F} \Vert_\mathcal{F} >0$ for~$f_\mathcal{F}\neq 0$. All functions~$f_\mathcal{F}$ must additionally be equipped with an inner-product in~$\mathcal{F}$. Simply speaking, a Hilbert space is an infinite dimensional vector space, where many operations behave like in the finite case. The properties of Hilbert spaces have been explored in great detail in literature, e.g., in~\cite{debnath2005introduction}. An extremely useful property of Hilbert spaces is that they are equivalent to an associated kernel function~\cite{aronszajn1950theory}. This equivalence allows to simply define a kernel, instead of fully defining the associated vector space. Formally speaking, if a Hilbert space~$\mathcal{H}$ is a RKHS, it will have a unique positive definite kernel~$k \colon \Z\times\Z\to\R$, which spans the space~$\mathcal{H}$. 
\begin{thm}[Moore-Aronszajn~\cite{aronszajn1950theory}] 
	Every positive definite kernel~$k$ is associated with a unique RKHS~$\mathcal{H}$.
\end{thm}
\begin{thm}[\cite{aronszajn1950theory}]
	Let~$\mathcal{F}$ be a Hilbert space,~$\Z$ a non-empty set and~$\bm{\phi}\colon\Z\to\mathcal{F}$. Then, the inner product~$\langle \bm{\phi}(\bm{z}),\bm{\phi}(\bm{z}^\prime)\rangle_\mathcal{F}\coloneqq k(\bm{z},\bm{z}^\prime)$ is positive definite.
\end{thm}
Importantly, any function~$f_\mathcal{H}$ in~$\mathcal{H}$ can be represented as a weighted linear sum of this kernel evaluated over the space~$\mathcal{H}$, as
\begin{align}
	f_\mathcal{H}(\cdot)=\langle f_\mathcal{H}(\cdot),k(x,\cdot)\rangle_{\mathcal{H}}=\sum_{i=1}^{n_\phi} \alpha_i k\left(\x_\text{dat}^{\{i\}},\cdot\right),
\end{align}
with~$\alpha_i\in\R$ for all~$i=\{1,\dots,n_\phi\}$, where~$n_\phi\in\N\cup\{\infty\}$ is the dimension of the feature space~$\mathcal{F}$. Thus, the RKHS is equipped with the inner-product
\begin{align}
	\langle f_\mathcal{H},f_\mathcal{H}^\prime\rangle_{\mathcal{H}}=\sum_{i=1}^{n_\phi}\sum_{j=1}^{n_\phi}\alpha_i\alpha_j^\prime k(\x_\text{dat}^{\{i\}},\x_\text{dat}^{\prime\{j\}}),
\end{align}
with~$f_\mathcal{H}^\prime(\cdot)=\sum_{j=1}^{n_\phi} \alpha_j^\prime k\left(\x_\text{dat}^{\prime\{j\}},\cdot\right)\in\mathcal{H},\alpha_j^\prime\in\R$. Now, the reproducing character manifests as
\begin{align}
	\forall \bm{z}\in\Z,\forall f_\mathcal{H}\in\mathcal{H},\,\langle f_\mathcal{H},k(x,\cdot)\rangle_{\mathcal{H}}=f_\mathcal{H}(\bm{z}),\text{ in particular }k(\bm{z},\bm{z}^\prime)=\langle k(\cdot,\bm{z}),k(\cdot,\bm{z}^\prime)\rangle_{\mathcal{H}}.\label{sec2:for:repk}
\end{align}
According to~\cite{steinwart2006explicit}, the RKHS is then defined as
\begin{align}
	\mathcal{H}=\{f_\mathcal{H}\colon \Z\to \R |\exists \bm{c}\in \mathcal{F},f_\mathcal{H}(\bm{z})=\langle \bm{c},\bm{\phi}(\bm{z})\rangle_{\mathcal{F}},\forall\bm{z}\in \Z\},\label{sec2:for:rkhs}
\end{align}
where~$\bm{\phi}(\bm{z})$ is the feature map constructing the kernel through~$k(\bm{z},\bm{z}^\prime)=\langle \bm{\phi}(\bm{z}),\bm{\phi}(\bm{z}^\prime)\rangle_\mathcal{F}$. 
\begin{exam}
		We want to find the RKHS for the polynomial kernel with degree~$2$ that is given by
	\begin{align*}
		k(\bm{z},\bm{z}^\prime)=(\bm{z}^\top \bm{z}^\prime)^2=(z_1 z_1^\prime)^2+2(z_1 z_1^\prime z_2 z_2^\prime)+(z_2 z_2^\prime)^2.
	\end{align*}
	for any~$\bm{z},\bm{z}^\prime\in\R^2$. First, we have to find a feature map~$\bm{\phi}$ such that the kernel corresponds to the inner product~$k(\bm{z},\y)=\langle \bm{\phi}(\bm{z}),\bm{\phi}(\y)\rangle$. A possible candidate for the feature map is 
	\begin{align*}
		\bm{\phi}(\bm{z})&=\begin{bmatrix}
		z_1^2,\sqrt{2}z_1 z_2,z_2^2
		\end{bmatrix}^\top\text{, because}\\
		\langle \bm{\phi}(\bm{z}),\bm{\phi}(\bm{z}^\prime)\rangle_{\R^3}&=\bm{\phi}(\bm{z})^\top\bm{\phi}(\y)=(z_1 z_1^\prime)^2+2(z_1 z_1^\prime z_2 z_2^\prime)+(z_2 z_2^\prime)^2=k(\bm{z},\bm{z}^\prime).
	\end{align*}
	We know that the RKHS contains all linear combinations of the form 
	\begin{align*}
		f_\mathcal{H}(\bm{z})&=\sum_{i=1}^3 \alpha_i k\left(\x_\text{dat}^{\{i\}},\bm{z}\right)=\sum_{i=1}^3 \alpha_i \langle \bm{\phi}(\bm{z}^\prime),\bm{\phi}(\bm{z})\rangle_{\R^3}=\sum_{i=1}^3 \langle \bm{c},\bm{\phi}(\bm{z})\rangle_{\R^3}\\
		&=c_1 z_1^2+c_2 \sqrt{2}z_1 z_2+c_3 z_2^2,
	\end{align*}
	with~$\bm\alpha,\bm{c},\x_\text{dat}^{\{i\}}\in\R^3$. Therefore, a possible candidate for the RKHS~$\mathcal{H}$ is given by
	\begin{align}
		\mathcal{H}=\left\lbrace f_\mathcal{H}\colon\R^2\to\R\vert f_\mathcal{H}(\bm{z})=c_1 z_1^2+c_2 \sqrt{2}z_1 z_2+c_3 z_2^2,\bm{c}\in\R^3\right\rbrace\label{sec2:for:ex_hilbert}
	\end{align}
	Next, it must be checked if the proposed Hilbert space is the related RKHS to the polynomial kernel with degree~$2$. This is achieved in two steps: i) Checking if the space is a Hilbert space and ii) confirming the reproducing property. First, we can easily proof that this is a Hilbert space rewriting~$f_\mathcal{H}(\bm{z})=\bm{z}^\top S \bm{z}$ with symmetric matrix~$S\in\R^{2\times 2}$ and using the fact that~$\mathcal{H}$ is euclidean and isomorphic to~$S$. Second, the condition for an RKHS must be fulfilled, i.e., the reproducing property~$f_\mathcal{H}(\bm{z})=\langle f_\mathcal{H}(\cdot),k(\cdot,\bm{z})\rangle_{\mathcal{H}}$. Since we can write
	\begin{align*}
		\langle f_\mathcal{H}(\cdot),k(\cdot,\bm{z})\rangle_{\mathcal{H}}=\langle \bm{c}^\top\bm{\phi}(\cdot),k(\cdot,\bm{z})\rangle_{\mathcal{H}}=\sum_{i=1}^3 c_i k(\cdot,\bm{z})= \bm{c}^\top\bm{\phi}(\bm{z})=f_\mathcal{H}(\bm{z}),
	\end{align*}
	property \cref{sec2:for:repk} is fulfilled and, thus,~$\mathcal{H}$ is the RKHS for the polynomial kernel with degree~$2$. Note that, even though the mapping~$\bm\phi$ is not unique for the kernel~$k$, the relation of~$k$ and the RKHS~$\mathcal{H}$ is unique.
\end{exam}
Given a function~$f_\mathcal{H}\in\mathcal{H}$ defined by~${n_\D}$ observations, its RKHS norm is defined as
\begin{align}
	\Vert f_\mathcal{H} \Vert_{\mathcal{H}}^2=\langle f_\mathcal{H},f_\mathcal{H}\rangle_{\mathcal{H}}=\sum_{i=1}^{{n_\D}}\sum_{j=1}^{{n_\D}}\alpha_i\alpha_j k(\x_\text{dat}^{\{i\}},\x_\text{dat}^{\prime\{j\}})=\bm{\alpha}^\top K(X,X) \bm{\alpha},\label{sec2:for:rkhs2}
\end{align}
with~$\bm{\alpha}\in\R^{n_\D}$ and~$K(X,X)$ given by~\cref{sec2:for:gp_K}. We can also use the feature map such that 
\begin{align}
	\Vert f_\mathcal{H} \Vert_{\mathcal{H}}=\inf\{\Vert \bm{c} \Vert_{\mathcal{F}}\colon \bm{c}\in \mathcal{F},f_\mathcal{H}(\bm{z})=\langle \bm{c},\bm{\phi}(\bm{z})\rangle_{\mathcal{F}},\forall \bm{z}\in \Z\}.\label{sec2:for:rkhs3}
\end{align}
As there is a unique relation between the RKHS~$\mathcal{H}$ and the kernel~$k$, the norm~$\Vert f_\mathcal{H} \Vert_{\mathcal{H}}$ can equivalently be written as~$\Vert f_\mathcal{H} \Vert_{k}$. The norm of a function in the RKHS indicates how fast the function varies over~$\Z$ with respect to the geometry defined by the kernel. Formally, it can be written as
\begin{align}
	\frac{\vert f_\mathcal{H}(\bm{z})-f_\mathcal{H}(\bm{z}^\prime)\vert}{d(\bm{z},\bm{z}^\prime)}\leq \Vert f_\mathcal{H} \Vert_{\mathcal{H}},\label{sec2:for:lip}
\end{align}
with the distance~$d(\bm{z},\bm{z}^\prime)^2=k(\bm{z},\bm{z})-2k(\bm{z},\bm{z}^\prime)+k(\bm{z}^\prime,\bm{z}^\prime)$. A function with finite RKHS norm is also element of the RKHS. A more detailed discussion about RKHS and norms is given in~\cite{wahba1990spline}.\newpage
\begin{exam}
	We want to find the RKHS norm of a function~$f_\mathcal{H}$ that is an element of the RKHS of the polynomial kernel with degree~$2$ that is given by
	\begin{align*}
		k(\bm{z},\bm{z}^\prime)=(\bm{z}^\top \bm{z}^\prime)^2=(z_1 z_1^\prime)^2+2(z_1 z_1^\prime z_2 z_2^\prime)+(z_2 z_2^\prime)^2.
	\end{align*}
	Let the function be
	\begin{align}
		f_\mathcal{H}(\bm{z})&=\sum_{i=1}^3 \alpha_i k\left(\x_\text{dat}^{\{i\}},\bm{z}\right),\text{ with}\label{sec2:for:examf}\\
		\alpha_1&=1,\,\alpha_2=-2,\,\alpha_3=3\label{sec2:for:examf1}\\
		\x_\text{dat}^{\{1\}}&=[1,1]^\top,\,\x_\text{dat}^{\{2\}}=[1,2]^\top,\,\x_\text{dat}^{\{3\}}=[2,1]^\top.\label{sec2:for:examf2}
	\end{align}
	Hence, function~\cref{sec2:for:examf} with~\cref{sec2:for:examf1,sec2:for:examf2} corresponds to
	\begin{align*}
		f_\mathcal{H}(\bm{z})&=11 z_1^2+6 z_1 z_2-4 z_2^2.
	\end{align*}
	Now, we have two possibilities how to calculate the RKHS norm. First, the RKHS-norm of~$f_\mathcal{H}$ is calculated using~\cref{sec2:for:rkhs2} by
	\begin{align*}
		\Vert f_\mathcal{H}\Vert_\mathcal{H}^2=\bm{\alpha}^\top K(X,X) \bm{\alpha}=
		\begin{bmatrix}
		1 & -2 & 3
		\end{bmatrix}\begin{bmatrix}
		4 & 9 & 9\\9 & 25 & 16\\9 & 16 & 25
		\end{bmatrix}\begin{bmatrix}
		1 \\ -2 \\ 3
		\end{bmatrix}
		=155
	\end{align*}
	with~$X=[\x_\text{dat}^{\{1\}},\x_\text{dat}^{\{2\}},\x_\text{dat}^{\{3\}}]$. Alternatively, we can use~\cref{sec2:for:rkhs3} that results in~$\Vert f_\mathcal{H}\Vert_\mathcal{H}=\Vert\bm{c}\Vert$, where~$\bm{c}$ is defined by~\cref{sec2:for:ex_hilbert}. Thus, the norm is computed as
	\begin{align*}
		f_\mathcal{H}(\bm{z})&=11 z_1^2+6 z_1 z_2-4 z_2^2\Rightarrow c_1=11,\,c_2=\frac{6}{\sqrt{2}},\,c_3=-4\Rightarrow \Vert f_\mathcal{H}\Vert_\mathcal{H}^2=155.
	\end{align*}
\end{exam}
\newpage
\begin{exam}
	In this example, we visualize the meaning of the RKHS norm.~\Cref{sec2:fig:rkhs_example} shows different quadratic functions with the same RKHS norm (top left and top right), a smaller RKHS norm (bottom left) and a larger RKHS norm (bottom right). An identical norm indicates a similar variation of the functions, whereas a higher norm leads to a more varying function.
		\begin{center}
			\tikzsetnextfilename{section2_rkhs_example}
			\captionsetup{type=figure}\begin{tikzpicture}
    \begin{axis}[	height=5cm,
    				width=5.8cm,
    				name=plot1,
    				view={-38}{30},
    				axis lines*=left, 
    				xlabel={$z_1$},
    				ylabel={$z_2$},
    				zlabel={$f_1(\bm z)$},
    				font={\sffamily},
    				xlabel shift=-5pt,
    				ylabel shift=-5pt,
    				zlabel shift=-5pt,
    				zmin=-33.2, zmax=18.4,
    				scale only axis]
\addplot3[surf,mesh/cols=21, mesh/rows=21] table {data/section2/rkhs_example1.dat};
\end{axis}
    \begin{axis}[	height=5cm,
    				width=5.8cm,
    				at=(plot1.right of south east), anchor=left of south west,
    				name=plot2,
    				view={-38}{30},
    				axis lines*=left, 
    				xlabel={$z_1$},
    				ylabel={$z_2$},
    				zlabel={$f_2(\bm z)$},
    				font={\sffamily},
    				xlabel shift=-5pt,
    				ylabel shift=-5pt,
    				zlabel shift=-5pt,
    				zmin=-33.2, zmax=18.4,
    				scale only axis]
\addplot3[surf,mesh/cols=21, mesh/rows=21] table {data/section2/rkhs_example2.dat};
\end{axis}
    \begin{axis}[	height=5cm,
    				width=5.8cm,
    				at=(plot1.below south east), anchor=above north east,
    				name=plot3,
    				view={-38}{30},
    				axis lines*=left, 
    				font={\sffamily},
    				xlabel={$z_1$},
    				ylabel={$z_2$},
    				zlabel={$f_3\bm z)$},
    				xlabel shift=-5pt,
    				ylabel shift=-5pt,
    				zlabel shift=-5pt,
    				zmin=-33.2, zmax=18.4,
    				scale only axis]
\addplot3[surf,mesh/cols=21, mesh/rows=21] table {data/section2/rkhs_example3.dat};
\end{axis}
    \begin{axis}[	height=5cm,
    				width=5.8cm,
    				at=(plot3.right of south east), anchor=left of south west,
    				name=plot4,
    				view={-38}{30},
    				axis lines*=left, 
    				font={\sffamily},
    				xlabel={$z_1$},
    				ylabel={$z_2$},
    				zlabel={$f_4(\bm z)$},
    				xlabel shift=-5pt,
    				ylabel shift=-5pt,
    				zlabel shift=-5pt,
    				zmin=-33.2, zmax=18.4,
    				scale only axis]
\addplot3[surf,mesh/cols=21, mesh/rows=21] table {data/section2/rkhs_example4.dat};
\end{axis}
\end{tikzpicture}
			\vspace{-0.0cm}
			
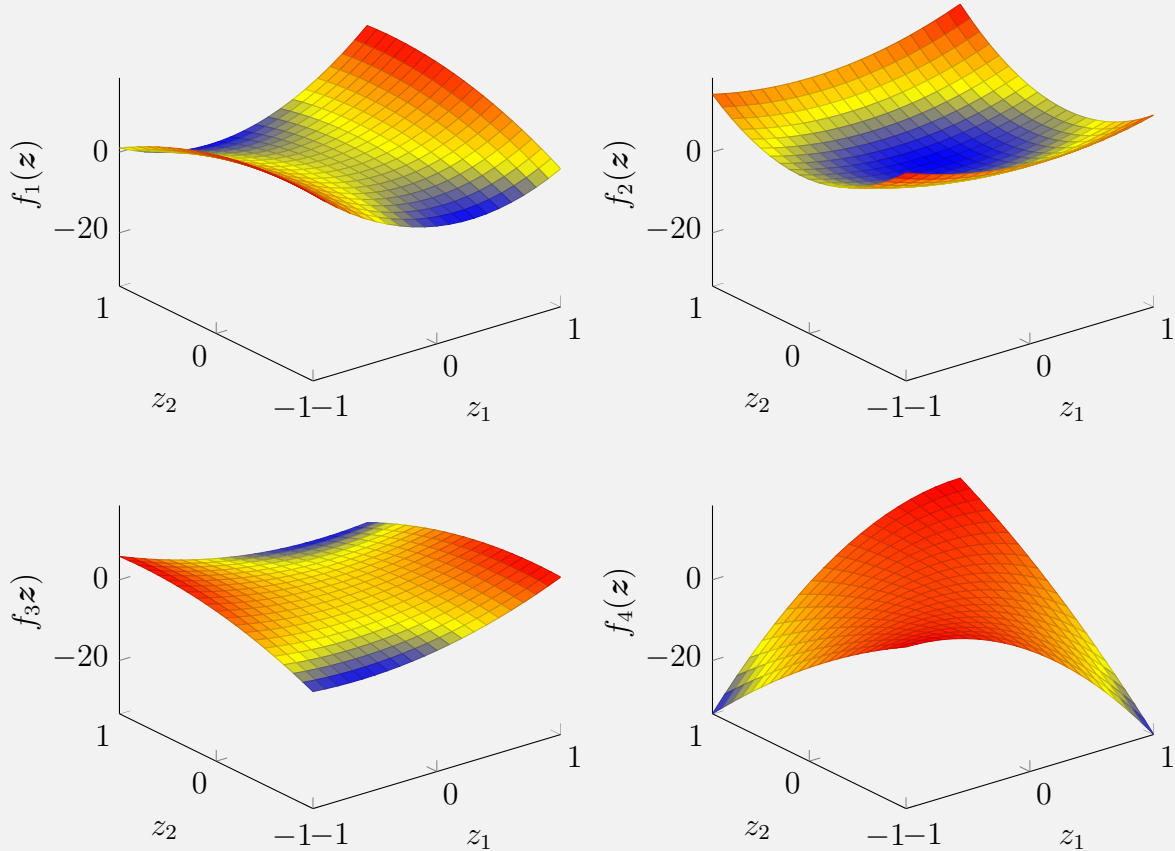
\captionof{figure}{Functions with different RKHS-norms:~$\Vert f_1 \Vert_\mathcal{H}^2\!=\!\Vert f_2 \Vert_\mathcal{H}^2\!=\!4\Vert f_3 \Vert_\mathcal{H}^2\!=\!\frac{1}{2}\Vert f_4 \Vert_\mathcal{H}^2$.}
			\label{sec2:fig:rkhs_example}
		\end{center}
\end{exam}
In summary, we investigate the unique relation between the kernel and its RKHS. The reproducing property allows us to write the inner-product as a tractable function which implicitly defines a higher (or even infinite) feature dimensional space. The RKHS-norm of a function is a Lipschitz-like indicator based on the metric defined by the kernel. This view of the RKHS is related to the kernel trick in machine learning. In the next section, the RKHS-norm is exploited to determine the error between the prediction of GPR and the actual data-generating function.
\subsection{Model Error}
\label{sec2:sec:modelerror}
One of the most interesting properties of GPR is the uncertainty description encoded in the predicted variance. This uncertainty is beneficial to quantify the error between the actual underlying data generating process and the GPR. In this section, we assume that there is an unknown function~$f_{\text{uk}}\colon\R^{n_z}\to\R$ that generates the training data. In detail, the data set~$\D=\{X,Y\}$ consists of
\begin{align}
	\begin{split}\label{sec2:for:medata}
		X&=[\x_\text{dat}^{\{1\}},\x_\text{dat}^{\{2\}},\ldots,\x_\text{dat}^{\{n_\D\}}]\in\R^{n_z\times {n_\D}}\\
		Y&=[\tilde y_\text{dat}^{\{1\}},\tilde y_\text{dat}^{\{2\}},\ldots,\tilde y_\text{dat}^{\{{n_\D}\}}]^\top\in\R^{{n_\D}},
	\end{split}
\end{align} 
where the data is generated by
\begin{align}
	\tilde y_\text{dat}^{\{i\}}=f_{\text{uk}}(\x_\text{dat}^{\{i\}})+\nu,\,\nu\sim\mathcal{N}(0,\sigma_n^2)
\end{align}
for all~$i=\{1,\ldots,{n_\D}\}$. Without any assumptions on~$f_{\text{uk}}$ it is obviously not possible to quantify the model error. Loosely speaking, the prior distribution of the GPR with kernel~$k$ must be suitable to learn the unknown function. More technically,~$f_{\text{uk}}$ must be an element of the RKHS spanned by the kernel as described in~\cref{sec2:for:rkhs}. This leads to the following assumption.
\begin{assum}\label{sec2:assum:rkhs}
	The function~$f_{\text{uk}}$ has a finite RKHS norm with respect to the kernel~$k$, i.e.,~$\Vert f_\text{uk} \Vert_{\mathcal{H}}<\infty$, where~$\mathcal{H}$ is the RKHS spanned by~$k$.
\end{assum}
This sounds paradoxical as~$f_{\text{uk}}$ is assumed to be unknown. However, there exist kernels that can approximate any continuous function arbitrarily exact. Thus, for any continuous function, an arbitrarily close function is element of the RKHS of an universal kernel. For more details, we refer to~\cref{sec2:sec:RKHS}. More infomation about the model error of misspecified Gaussian Process model can be found in~\cite{beckers:cdc2018}\\
We classify the error quantification in three different approaches: i) the robust approach, ii) the scenario approach, and iii) the information-theoretical approach. The different techniques are presented in the following and visualized in~\cref{sec2:fig:model_error}. For the remainder of this section, we assume that a GPR is trained with the data set~\cref{sec2:for:medata} and~\cref{sec2:assum:rkhs} holds.
\subsubsection{Robust approach}
The robust approach exploits the fact that the prediction of the GPR is Gaussian distributed. Thus, for any~$\bm{z}^*\in\R^{n_z}$, the model error is bounded by
\begin{align}
	\vert f_{\text{uk}}(\bm{z}^*)- \mean(f_\text{GP}\vert \bm{z}^*,\mathcal D)\vert\leq c\var(f_\text{GP}\vert \bm{z}^*,\mathcal D)
\end{align}
with high probability where~$c\in\R_{>0}$ adjusts the probability. However, for multiple test points~$\bm{z}^*_1,\bm{z}^*_2,\ldots\in\R^{n_z}$, this approach neglects any correlation between~$f_\text{GP}(\bm{z}^*_1),f_\text{GP}(\bm{z}^*_2),\ldots$. \Cref{sec2:fig:model_error} shows how for a given~$\bm{z}^*_1$ and~$\bm{z}^*_2$, the variance is exploited as upper bound. Thus, any prediction is handled independently, which leads to a very conservative bound, see~\cite{umlauft:ecc2018}. 
\subsubsection{Scenario approach}
Instead of using the mean and the variance as in the robust approach, the scenario approach deals with the samples of the GPR directly. In contrast to the other methods, there is no direct model error quantification but rather a sample based quantification. The idea is to draw a large number~$n_\text{scen}\in\N$ of sample functions~$f_\text{GP}^1,f_\text{GP}^2,\ldots,f_\text{GP}^{n_\text{scen}}$ over $n_s\in\N$ sampling points. The sampling is performed by drawing multiple instances from~$f_\text{GP}$ given by the multivariate Gaussian distribution
\begin{align}
	\begin{bmatrix} Y\vphantom{\begin{bmatrix}m(\x_\text{dat}^{\{1\}})\\\vdots\\ m(\x_\text{dat}^{\{{n_\D}\}})\end{bmatrix}} \\ f_\text{GP}(\bm{z}^*_1)\\\vdots \\ f_\text{GP}(\bm{z}^*_{n_s}) \end{bmatrix}\sim \mathcal{N} \left(\begin{bmatrix}m(\x_\text{dat}^{\{1\}})\\\vdots\\ m(\x_\text{dat}^{\{{n_\D}\}})\\m(\bm{z}^*_1)\\\vdots\\m(\bm{z}^*_{n_s})\end{bmatrix}, \begin{bmatrix} K(X,X)+\sigma_n^2 I_{n_\D}\vphantom{\begin{bmatrix}m(\x_\text{dat}^{\{1\}})\\\vdots\\ m(\x_\text{dat}^{\{{n_\D}\}})\end{bmatrix}} & K(X^*,X)\\ K(X^*,X)^\top \vphantom{\begin{bmatrix}m(\x_\text{dat}^{\{1\}})\\\vdots\\ m(\x_\text{dat}^{\{{n_\D}\}})\end{bmatrix}} & K(X^*,X^*) \end{bmatrix}\right),\label{sec2:for:scenapp}
\end{align} 
where~$X^*=[\bm{z}^*_1,\cdots,\bm{z}^*_{n_s}]$ contains the sampling points. Each sample can then be used in the application instead of the unknown function. For a large number of samples it is assumed that the unknown function is close to one of these samples. However, the crux of this approach is to determine, for a given model error~$c\in\R_{>0}$, the required number of samples~$n_\text{scen}$ and probability~$\delta_{scen}>0$ such that
\begin{align}
\Prob\big(\vert f_{\text{uk}}(\bm{z}^*)- f_\text{GP}^i(\bm{z}^*)\vert \leq c,i\in\{1,\ldots,n_\text{scen}\}\big)\geq \delta_{scen}
\end{align}
for all~$\bm{z}^*\in Z$. In \cref{sec2:fig:model_error}, five different samples of a GP model are drawn as example.
\subsubsection{Information-theoretical approach}
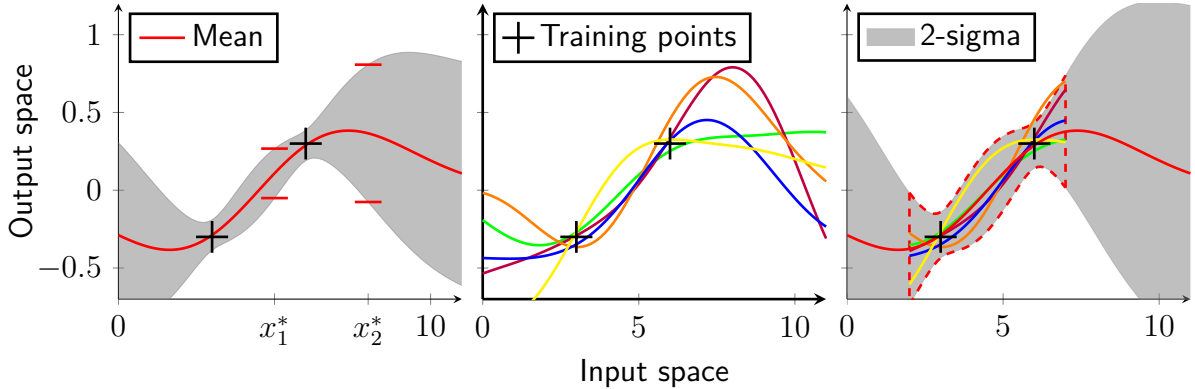
\begin{figure}
		\tikzsetnextfilename{section2_model_error}
		\begin{tikzpicture}
	\begin{axis}[
	  name=plot1,
	  axis lines=left,
	  ylabel={Output space},
	  line width=1pt,
	  grid = none,
	  height=5.5cm,
	  width=6.1cm,
	  legend pos={north west},
	  xtick={0,5,8,10},
	  font={\sffamily},
	  xticklabels={$0$,$x^*_1$,$x^*_2$,$10$},
	  xmin=0, xmax=11,ymin=-0.7,ymax=1.2,
	 ylabel style={at={(-0.2,0.5)}}]
			 \addplot[name path=varp1, color=gray,opacity=0.5, no marks] table [x index=0,y expr=\thisrowno{5}+2*(\thisrowno{6})]{data/section2/figure3_var.dat};
			 \addplot[name path=varm1, color=gray,opacity=0.5, no marks] table [x index=0,y expr=\thisrowno{5}-2*(\thisrowno{6})]{data/section2/figure3_var.dat};
			 \addplot[gray,opacity=0.5] fill between[ of = varm1 and varp1]; 
			 \addplot[mark=+,color=black, only marks,mark size=6,line width=1pt,select coords between index={0}{2}] table [x index=0,y index=1]{data/section2/figure3_data.dat};
			 \addplot[mark=-,color=red, only marks,mark size=5,line width=1pt,select coords between index={72}{72}] table [x index=0,y expr=\thisrowno{5}+2*(\thisrowno{6})]{data/section2/figure3_var.dat};
			 \addplot[mark=-,color=red, only marks,mark size=5,line width=1pt,select coords between index={72}{72}] table [x index=0,y expr=\thisrowno{5}-2*(\thisrowno{6})]{data/section2/figure3_var.dat};
			 \addplot[mark=-,color=red, only marks,mark size=5,line width=1pt,select coords between index={45}{45}] table [x index=0,y expr=\thisrowno{5}+2*(\thisrowno{6})]{data/section2/figure3_var.dat};
			 \addplot[mark=-,color=red, only marks,mark size=5,line width=1pt,select coords between index={45}{45}] table [x index=0,y expr=\thisrowno{5}-2*(\thisrowno{6})]{data/section2/figure3_var.dat};
  			 \addplot[color=red,line width=1pt] table [x index=0,y index=5]{data/section2/figure3_var.dat};	 
  			 \legend{,,,,,,,,Mean}
	\end{axis}
	\begin{axis}[
	  name=plot2,
	  xshift=-1mm,
	  at=(plot1.right of south east), anchor=left of south west,
	  axis lines=left,
	  xlabel={Input space},
	  font={\sffamily},
	  line width=1pt,
	  grid = none,
	  legend pos={north west},
	  height=5.5cm,
	  width=6.1cm,
	  yticklabels={,,},
	  xmin=0, xmax=11,ymin=-0.7,ymax=1.2]
			 \addplot[mark=+,color=black, only marks,mark size=6,line width=1pt,select coords between index={0}{2}] table [x index=0,y index=1]{data/section2/figure3_data.dat};
  			 \addplot[color=green,line width=1pt] table [x index=0,y index=11]{data/section2/figure3_samples.dat};
			 \addplot[color=purple,line width=1pt] table [x index=0,y index=12]{data/section2/figure3_samples.dat};
			 \addplot[color=orange,line width=1pt] table [x index=0,y index=13]{data/section2/figure3_samples.dat};
			 \addplot[color=blue,line width=1pt] table [x index=0,y index=14]{data/section2/figure3_samples.dat};
			 \addplot[color=yellow,line width=1pt] table [x index=0,y index=15]{data/section2/figure3_samples.dat};	 
			 \legend{Training points}
	\end{axis}
		\begin{axis}[
	  name=plot3,
	  at=(plot2.right of south east), anchor=left of south west,
	  xshift=-1mm,
	  axis lines=left,
	  font={\sffamily},
	  line width=1pt,
	  grid = none,
	  height=5.5cm,
	  width=6.1cm,
	  yticklabels={,,},
	  legend pos={north west},
	  xmin=0, xmax=11,ymin=-0.7,ymax=1.2]
			 \addplot[name path=varp1, color=gray,opacity=0.3] table [x index=0,y expr=\thisrowno{5}+3*(\thisrowno{6})]{data/section2/figure3_var.dat};
			 \addplot[name path=varm1, color=gray,opacity=0.3] table [x index=0,y expr=\thisrowno{5}-3*(\thisrowno{6})]{data/section2/figure3_var.dat};
			 \addplot[name path=varp2, color=red, dashed,line width=1pt, no marks,select coords between index={18}{63}] table [x index=0,y expr=\thisrowno{5}+3*(\thisrowno{6})]{data/section2/figure3_var.dat};
			 \addplot[name path=varm2, color=red, dashed,line width=1pt, no marks,select coords between index={18}{63}] table [x index=0,y expr=\thisrowno{5}-3*(\thisrowno{6})]{data/section2/figure3_var.dat};
			 \addplot[color=red, dashed,line width=1pt, no marks] coordinates {(2,-0.01595679) (2,-0.73765107)};
			 \addplot[color=red, dashed,line width=1pt, no marks] coordinates {(7,0.01595679) (7,0.73765107)};
			 \addplot[gray,opacity=0.5] fill between[ of = varm1 and varp1]; 
			 \addplot[mark=+,color=black, only marks,mark size=6,line width=1pt,select coords between index={0}{2}] table [x index=0,y index=1]{data/section2/figure3_data.dat};
  			 \addplot[color=green,line width=1pt,select coords between index={18}{63}] table [x index=0,y index=11]{data/section2/figure3_samples.dat};
			 \addplot[color=purple,line width=1pt,select coords between index={18}{63}] table [x index=0,y index=12]{data/section2/figure3_samples.dat};
			 \addplot[color=orange,line width=1pt,select coords between index={18}{63}] table [x index=0,y index=13]{data/section2/figure3_samples.dat};
			 \addplot[color=blue,line width=1pt,select coords between index={18}{63}] table [x index=0,y index=14]{data/section2/figure3_samples.dat};
			 \addplot[color=yellow,line width=1pt,select coords between index={18}{63}] table [x index=0,y index=15]{data/section2/figure3_samples.dat};	
  			 \addplot[color=red,line width=1pt] table [x index=0,y index=5]{data/section2/figure3_var.dat};	
  			 \legend{,,,,,,2-sigma} 
	\end{axis}
	\end{tikzpicture} 
		\vspace{-0.2cm}\caption{Different approaches to quantify the model error: Robust approach (left), scenario approach (middle), information-theoretical approach (right).}\vspace{-0.2cm}
		\label{sec2:fig:model_error}
	\end{figure}

Alternatively, the work in~\cite{srinivas2012information} derives an upper bound for samples of the GPR on a compact set with a specific probability. In contrast to the robust approach, the correlation between the function values are considered. We restate here the theorem of~\cite{srinivas2012information}.
\begin{thm}[\cite{srinivas2012information}]\label{sec2:thm:srin}
Given~\cref{sec2:assum:rkhs}, the model error~$\Delta\in\R$
	\begin{align}
		\Delta=\vert\mean(f_\text{GP}\vert \bm{z},\mathcal D)- f_{\text{uk}}(\bm{z})\vert
	\end{align}
	is bounded for all $\bm{z}$ on a compact set $\Omega\subset\R^{n_z}$ with a probability of at least~$\delta\in(0,1)$ by
	\begin{align}
		\text{P}&\left\lbrace \forall \bm{z}\in \Omega,\, \Delta\leq \vert \beta \Var^{\frac{1}{2}}(f_\text{GP}\vert \bm{z},\mathcal D)\vert\right\rbrace\geq \delta,\label{sec2:for:probbound}
	\end{align}
	where~$\beta\in\R$ is defined as
	\begin{align}
		\beta&=\sqrt{2\Verts{f_{\text{uk}}}^2_k+300 \gamma_{\text{max}} \ln^3\left(\frac{{n_\D}+1}{1-\delta}\right)}.
	\end{align}
	The variable~$\gamma_{\text{max}}\in\R$ is the maximum of the information gain 
	\begin{align}
		\gamma_{\text{max}}&= \hspace{-0.2cm}\max_{\x_\text{dat}^{\{1\}},\ldots,\x_\text{dat}^{\{{n_\D}+1\}}\in \Omega}\frac{1}{2}\log \vert I_{n_\D+1}+\sigma_n^{-2}K(\bm{z},\bm{z}^\prime)\vert
	\end{align}
	with Gram matrix~$K(\bm{z},\bm{z}^\prime)$ and the input elements~$\bm{z},\bm{z}^\prime\in\{\x_\text{dat}^{\{1\}},\ldots,\x_\text{dat}^{\{{n_\D}+1\}}\}$.
\end{thm}
To compute this bound, the RKHS norm of~$f_{\text{uk}}$ must be known. That is in application usually not the case. However, often the norm can be upper bounded and thus, the bound in~\cref{sec2:thm:srin} can be upper bounded. For this purpose, the relation of the RKHS norm to the Lipschitz constant given by~\cref{sec2:for:lip} is beneficial as the Lipschitz constant is more likely to be known. In general, the computation of the information gain is a non-convex optimization problem. However, the information capacity~$\gamma_{\text{max}}$ has a sub-linear dependency on the number of training points for many commonly used kernel functions~\cite{srinivas2012information}. Therefore, even though~$\beta$ is increasing with the number of training data, it is possible to learn the true function~$f_{\text{uk}}$ arbitrarily exactly~\cite{Berkenkamp2016ROA}. In contrast to the other approaches, \cref{sec2:thm:srin} allows to bound the error for any test point in a compact set. In \cite{beckers2019automatica}, we exploit this approach in GP model based control tasks. The right illustration of~\cref{sec2:fig:model_error} visualizes the information-theoretical bound.
\section{Model Selection}
\label{sec2:sec:kernel}
\Cref{sec2:for:gp_post} clearly shows the immense impact of the kernel on the posterior mean and variance. However, this is not surprising as the kernel is an essential part of the prior model. For practical applications that leads to the question how to choose the kernel. Additionally, most kernels depend on a set of hyperparameters that must be defined. Thus in order to turn GPR into a powerful practical tool it is essential to develop methods that address the model selection problem. We see the model selection as the determination of the kernel and its hyperparameters. We only focus on kernels that are defined on~$\Z\subseteq\R^{n_z}$. In the next two subsections, we present different kernels and explain the role of the hyperparameters and their selection, mainly based on~\cite{rasmussen2006gaussian}.
\begin{rem}
The selection of the kernel functions seems to be similar to the model selection for parametric models. However, there are two major differences: i) the selection is fully covered by the Bayesian methodology and ii) many kernels allow to model a wide range of different functions whereas parametric models a typically limited to very specific types of functions.
\end{rem}
\subsection{Kernel Functions}
The value of the kernel function~$k(\bm{z},\bm{z}^\prime)$ is an indicator of the interaction of two states~$(\bm{z},\bm{z}^\prime)$. Thus, an essential part of GPR is the selection of the kernel function and the estimation of its free parameters~$\varphi_1,\varphi_2,\ldots,\varphi_{n_\varphi}$, called hyperparameters. The number~$n_\varphi$ of hyperparameters depends on the kernel function. The choice of the kernel function and the determination of the corresponding hyperparameters can be seen as degrees of freedom of the regression. First of all, we start with the general properties of a function to be qualified as a kernel for GPR. A necessary and sufficient condition for the function~$k\colon \Z\times\Z\to\R$ to be a valid kernel is that the Gram matrix, see~\cref{sec2:for:gp_K}, is positive semidefinite for all possible input values~\cite{shawe2004kernel}.
\begin{rem}
	As shown in \cref{sec2:sec:RKHS}, the kernel function must be \emph{positive definite} to span an unique RKHS. That seems to be contradictory to the required \emph{positive semi-definiteness} of the Gram matrix. The solution is the definition of positive definite kernels as it is equivalent to a positive semi-definite Gram matrix. In detail, a symmetric function~$k\colon \Z\times\Z\to\R$ is a \emph{positive definite} kernel on~$\Z$ if
	\begin{align}
		\sum_{j=1}^{n_\D}\sum_{i=1}^{n_\D} k(\x_\text{dat}^{\{i\}},\x_\text{dat}^{\{j\}})c_i c_j\geq 0
	\end{align}
	holds for any~${n_\D}\in\N$,~$\x_\text{dat}^{\{1\}},\ldots,\x_\text{dat}^{\{{n_\D}\}}\in\Z$ and~$c_1,\ldots,c_n\in\R$. Thus, there exists a \emph{positive semi-definite} matrix~$A_G\in\R^{{n_\D}\times {n_\D}}$ such that
	\begin{align}
		\x_\text{dat}^\top A_G \x_\text{dat}=\sum_{j=1}^{n_\D}\sum_{i=1}^{n_\D} k(\x_\text{dat}^{\{i\}},\x_\text{dat}^{\{j\}})c_ic_j
	\end{align}
	holds for any~${n_\D}\in\N$ and~$\bm{z}\in\Z$.
\end{rem}
The set of functions~$k$ which fulfill this condition is denoted with~$\mathcal{K}$. Kernel functions can be separated into two classes, the \emph{stationary} and \emph{non-stationary} kernels. A stationary kernel is a function of the distance~$\bm{z}-\bm{z}^\prime$. Thus, it is invariant to translations in the input space. In contrast, non-stationary kernels depend directly on~$\bm{z}$,$\bm{z}^\prime$ and are often functions of a dot product~$\bm{z}^\top\bm{z}$. In the following, we list some common kernel functions with their basic properties. Even though the number of presented kernels is limited, new kernels can be constructed easily as~$\mathcal{K}$ is closed under specific operations such as addition and scalar multiplication. At the end, we summarize the equation of each kernel in~\cref{tab:kernel} and provide a comparative example.
\subsubsection{Constant Kernel}
The equation for the constant kernel is given by
	\begin{align}
		k(\bm{z},\bm{z}^\prime)=\varphi_1^2.\label{sec2:for:cvconst}
	\end{align}
	This kernel is mostly used in addition to other kernel functions. It depends one a single hyperparameter~$\varphi_1\in\R_{\geq 0}$.
\subsubsection{Linear Kernel}
The equation for the linear kernel is given by
	\begin{align}
		k(\bm{z},\bm{z}^\prime)=\bm{z}^\top \bm{z}^\prime.\label{sec2:for:covlin}
	\end{align}
	The linear kernel is a dot-product kernel and thus, non-stationary. The kernel can be obtained from Bayesian linear regression as shown in~\cref{sec2:sec:kbv}. The linear kernel is often used in combination with the constant kernel to include a bias.
\subsubsection{Polynomial Kernel}
The equation for the polynomial kernel is given by
	\begin{align}
		k(\bm{z},\bm{z}^\prime)=\left(\bm{z}^\top \bm{z}^\prime+ \varphi_1^2 \right)^p,\,p\in\N.\label{sec2:for:covpoly}
	\end{align}
	The polynomial kernel has an additional parameter~$p\in\N$, that determines the degree of the polynomial. Since a dot product is contained, the kernel is also non-stationary. The prior variance grows rapidly for~$\Vert \bm{z} \Vert>1$ such that the usage for some regression problems is limited. It depends on a single hyperparameter~$\varphi_1\in\R_{\geq 0}$.
\subsubsection{Matérn Kernel}
The equation for the Matérn kernel is given by
	\begin{align}
		k(\bm{z},\bm{z}^\prime)\!=\!\varphi_1^2\exp \!\left(-{\frac {{\sqrt {2\check{p} }}\Vert \bm{z}- \bm{z}^\prime \Vert}{\varphi_2}}\right)\!{\frac {p!}{(2p)!}}\sum _{i=0}^{p}{\frac {(p+i)!}{i!(p-i)!}}\!\left({\frac {{\sqrt {8\check{p} }}\Vert \bm{z}- \bm{z}^\prime \Vert}{\varphi_2}}\right)^{p-i}\label{sec2:for:covmatern}
	\end{align}
	with~$\check{p}=p+\frac{1}{2},p\in\N$. The Matérn kernel is a very powerful kernel and presented here with the most common parameterization for~$\check{p}$. Functions drawn from a GP model with Matérn kernel are~$p$-times differentiable. The more general equation of this stationary kernel can be found in~\cite{bishop2006pattern}. This kernel is an \emph{universal kernel} which is explained in the following.
	\begin{lem}[{\cite[Lemma 4.55]{steinwart2008support}}]\label{sec2:lem:uni}
		Consider the RKHS~$\mathcal{H}(\Z_c)$ of an universal kernel on any prescribed compact subset~$\Z_c\in\Z$. Given any positive number~$\varepsilon$ and any function~$f_\mathcal{C}\in\mathcal{C}^1(\Z_C)$, there is a function~$f_\mathcal{H}\in\mathcal{H}(\Z_c)$ such that~$\Vert f_\mathcal{C}-f_\mathcal{H}\Vert_{\Z_c}\leq \varepsilon$.
	\end{lem}
	Intuitively speaking, a GPR with an universal kernel can approximate any continuous function arbitrarily exact on a compact set. For~$p\to\infty$, it results in the squared exponential kernel. The two hyperparameters are~$\varphi_1\in\R_{\geq 0}$ and~$\varphi_2\in\R_{>0}$.
\subsubsection{Squared Exponential Kernel}
The equation for the squared exponential kernel is given by
	\begin{align}
		k(\bm{z},\bm{z}^\prime)=\varphi_1^2 \exp{\left(-\frac{\Vert \bm{z}- \bm{z}^\prime \Vert^2}{2\varphi_2^2} \right) }.\label{sec2:for:covse}
	\end{align}
	Probably the most widely used kernel function for GPR is the squared exponential kernel, see~\cite{rasmussen2006gaussian}. The hyperparameter~$\varphi_1$ describes the signal variance which determines the average distance of the data-generating function from its mean. The lengthscale~$\varphi_2$ defines how far it is needed to move along a particular axis in input space for the function values to become uncorrelated. Formally, the lengthscale determines the number of expected upcrossings of the level zero in a unit interval by a zero-mean GP.  The squared exponential kernel is infinitely differentiable, which means that the GPR exhibits a smooth behavior. As limit of the Matérn kernel, it is also an universal kernel, see~\cite{micchelli2006universal}. 
\begin{exam}
	\Cref{fig:flex_reg} shows the power for regression of universal kernel functions. In this example, a GPR with squared exponential kernel is used for different training data sets. The hyperparameter are optimized individually for each training data set by means of the likelihood, see~\cref{sec2:sec:hypopt}. Note that all presented regressions are based on the \textbf{same} GP model, i.e. the same kernel function, but with different data sets. That highlights again the superior flexibility of GPR.
	\begin{center}
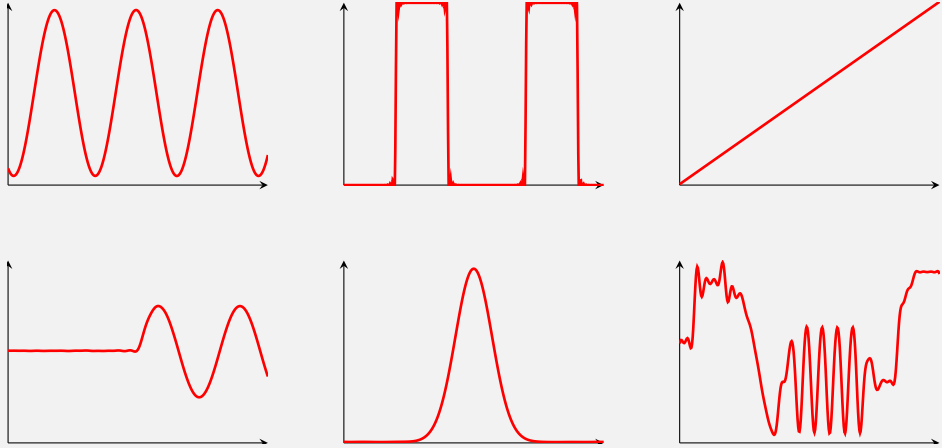

		\tikzsetnextfilename{section2_flex_reg}
		\captionsetup{type=figure}	\begin{tikzpicture}
\begin{groupplot}[
  group style={
    group size=3 by 2,
    group name=G},
 	  axis lines=left,
	  grid = none,
	  height=4cm,
	  width=5cm,
	  ticks=none,
	  xticklabels={,,},yticklabels={,,},
	  xmin=0, xmax=10,ymin=0,ymax=1]
\nextgroupplot
\addplot[color=red,no marks,line width=1pt] table [x index=0,y index=1]{data/section2/figure5_mean.dat};
\nextgroupplot
\addplot[color=red,no marks,line width=1pt] table [x index=0,y index=2]{data/section2/figure5_mean.dat};
\nextgroupplot
\addplot[color=red,no marks,line width=1pt] table [x index=0,y index=3]{data/section2/figure5_mean.dat};
\nextgroupplot
\addplot[color=red,no marks,line width=1pt] table [x index=0,y index=4]{data/section2/figure5_mean.dat};
\nextgroupplot
\addplot[color=red,no marks,line width=1pt] table [x index=0,y index=5]{data/section2/figure5_mean.dat};
\nextgroupplot
\addplot[color=red,no marks,line width=1pt] table [x index=0,y index=6]{data/section2/figure5_mean.dat};
\end{groupplot}
\end{tikzpicture} 
		\captionof{figure}{Examples for the flexibility of the regression that all are based on the same GP model.\label{fig:flex_reg}}
	\end{center}
\end{exam}
\subsubsection{Rational Quadratic Kernel}
The equation for the rational quadratic kernel is given by
	\begin{align}
		k(\bm{z},\bm{z}^\prime)=\varphi_1^2\left( 1+\frac{\Verts{\bm{z}-\bm{z}^\prime}^2}{2p \varphi_2^2}\right)^{-p},\,p\in\N.\label{sec2:for:covrq}
	\end{align}
	This kernel is equivalent to summing over infinitely many squared exponential kernels with different lengthscales. Hence, GP priors with this kernel are expected to see functions which vary smoothly across many lengthscales. The parameter~$p$ determines the relative weighting of large-scale and small-scale variations. For~$p\to\infty$, the rational quadratic kernel is identical to the squared exponential kernel.
\subsubsection{Squared Exponential ARD Kernel}
The equation for the squared exponential ARD kernel is given by
\begin{align}
	k(\bm{z},\bm{z}^\prime)=\varphi_1^2 \exp{\left(-(\bm{z}-\bm{z}^\prime)^\top P^{-1}(\bm{z}-\bm{z}^\prime) \right) },\, P=\diag(\varphi_2^2,\ldots,\varphi_{1+n_z}^2).\label{sec2:for:covseard}
\end{align}
The \emph{automatic relevance determination} (ARD) extension to the squared exponential kernel allows for independent lenghtscales~$\varphi_2,\ldots,\varphi_{1+n_z}\in\R_{>0}$ for each dimension of~$\bm{z},\bm{z}^\prime\in\R^{n_z}$. The individual lenghtscales are typically larger for dimensions which are irrelevant as the covariance will become almost independent of that input. A more detailed discussion about the advantages of different kernels can be found, for instance, in~\cite{mackay1997gaussian} and~\cite{bishop2006pattern}.
\begin{exam}
In this example, we use three GPRs with the same set of training data
\begin{align}
X=[1,3,5,7,9],\,Y=[0,1,2,3,6]
\end{align}
but with different kernels, namely the squared exponential~\cref{sec2:for:covse}, the linear~\cref{sec2:for:covlin}, and the polynomial~\cref{sec2:for:covpoly} kernel.~\Cref{sec2:fig:model_selection} shows the different shapes of the regressions with the posterior mean (red), the posterior variance (gray shaded) and the training points (black). Even for this simple data set, the flexibility of the squared exponential kernel is already visible.
	\begin{center}
		\tikzsetnextfilename{section2_model_selection}
		\captionsetup{type=figure}\begin{tikzpicture}
	\begin{axis}[
	  name=plot1,
	  axis lines=left,
	  ylabel={Output space},
	  font={\sffamily},
	  line width=1pt,
	  grid = none,
	  height=5cm,
	  width=6.1cm,
	  xmin=0, xmax=11,ymin=-2,ymax=8,
	 ylabel style={at={(-0.2,0.5)}}]
			 \addplot[name path=varp1, color=gray,opacity=0.3, no marks] table [x index=0,y expr=\thisrowno{1}+2*(\thisrowno{2})]{data/section2/model_selection_example_mean_var.dat};
			 \addplot[name path=varm1, color=gray,opacity=0.3, no marks] table [x index=0,y expr=\thisrowno{1}-2*(\thisrowno{2})]{data/section2/model_selection_example_mean_var.dat};
			 \addplot[gray,opacity=0.5] fill between[ of = varm1 and varp1]; 
			 \addplot[mark=+,color=black, only marks,mark size=6,line width=1pt] table [x index=0,y index=1]{data/section2/model_selection_example_data.dat};
  			 \addplot[color=red,line width=1pt] table [x index=0,y index=1]{data/section2/model_selection_example_mean_var.dat};	 
	\end{axis}
	\begin{axis}[
	  name=plot2,
	  xshift=-1mm,
	  at=(plot1.right of south east), anchor=left of south west,
	  axis lines=left,
	  font={\sffamily},
	  xlabel={Input space},
	  line width=1pt,
	  grid = none,
	  height=5cm,
	  width=6.1cm,
	  yticklabels={,,},
	  xmin=0, xmax=11,ymin=-2,ymax=8]
			 \addplot[name path=varp1, color=gray,opacity=0.3, no marks] table [x index=0,y expr=\thisrowno{3}+2*(\thisrowno{4})]{data/section2/model_selection_example_mean_var.dat};
			 \addplot[name path=varm1, color=gray,opacity=0.3, no marks] table [x index=0,y expr=\thisrowno{3}-2*(\thisrowno{4})]{data/section2/model_selection_example_mean_var.dat};
			 \addplot[gray,opacity=0.5] fill between[ of = varm1 and varp1]; 
			 \addplot[mark=+,color=black, only marks,mark size=6,line width=1pt] table [x index=0,y index=1]{data/section2/model_selection_example_data.dat};
  			 \addplot[color=red,line width=1pt] table [x index=0,y index=3]{data/section2/model_selection_example_mean_var.dat};		 
	\end{axis}
		\begin{axis}[
	  name=plot3,
	  at=(plot2.right of south east), anchor=left of south west,
	  xshift=-1mm,
	  axis lines=left,
	  font={\sffamily},
	  line width=1pt,
	  grid = none,
	  height=5cm,
	  width=6.1cm,
	  yticklabels={,,},
	  xmin=0, xmax=11,ymin=-2,ymax=8]
			 \addplot[name path=varp1, color=gray,opacity=0.3, no marks] table [x index=0,y expr=\thisrowno{5}+2*(\thisrowno{6})]{data/section2/model_selection_example_mean_var.dat};
			 \addplot[name path=varm1, color=gray,opacity=0.3, no marks] table [x index=0,y expr=\thisrowno{5}-2*(\thisrowno{6})]{data/section2/model_selection_example_mean_var.dat};
			 \addplot[gray,opacity=0.5] fill between[ of = varm1 and varp1]; 
			 \addplot[mark=+,color=black, only marks,mark size=6,line width=1pt] table [x index=0,y index=1]{data/section2/model_selection_example_data.dat};
  			 \addplot[color=red,line width=1pt] table [x index=0,y index=5]{data/section2/model_selection_example_mean_var.dat};	 
	\end{axis}
	\end{tikzpicture} 
		\vspace{-0.5cm}
		
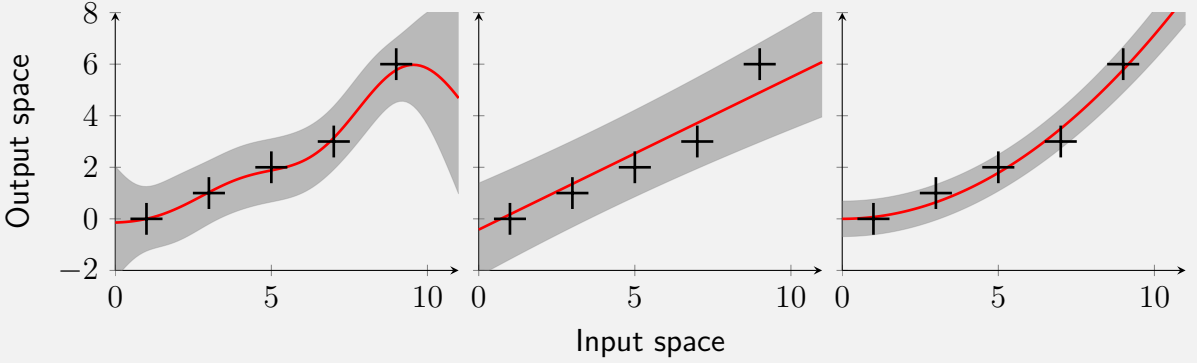
\captionof{figure}{GPR with different kernels: squared exponential (left), linear (middle) and polynomial with degree 2 (right).}
		\label{sec2:fig:model_selection}
	\end{center}
\end{exam}
\begin{table}[h]
\begin{center}
\begin{tabular}{ll}
\toprule
Kernel name & $k(\bm{z},\bm{z}^\prime)=$\\
\midrule
Constant & $\varphi_1^2$ \\ 
\addlinespace[0.4cm]
Linear & $\bm{z}^\top \bm{z}^\prime+ \varphi_1^2$\\ 
\addlinespace[0.4cm]
Polynomial~$p\in\N$ & $\left(\bm{z}^\top \bm{z}^\prime+ \varphi_1^2 \right)^p$\\ 
\addlinespace[0.4cm]
Matérn~$\check{p}=p+\frac{1}{2},p\in\N$ &~$\varphi_1^2\exp \left(-{\frac {{\sqrt {2\check{p} }}\Vert \bm{z}- \bm{z}^\prime \Vert}{\varphi_2}}\right){\frac {p!}{(2p)!}}\sum _{i=0}^{p}{\frac {(p+i)!}{i!(p-i)!}}\left({\frac {{\sqrt {8\check{p} }}\Vert \bm{z}- \bm{z}^\prime \Vert}{\varphi_2}}\right)^{p-i}$\\
\addlinespace[0.4cm]
Squared exponential& $\varphi_1^2 \exp{\left(-\frac{\Vert \bm{z}- \bm{z}^\prime \Vert^2}{2\varphi_2^2} \right) }$\\
\addlinespace[0.4cm]
Rational quadratic & $\varphi_1^2\left( 1+\frac{\Verts{\bm{z}-\bm{z}^\prime}^2}{2p \varphi_2^2}\right)^{-p}$\\
\addlinespace[0.4cm]
Squared exponential ARD & $\varphi_1^2 \exp{\left(-(\bm{z}-\bm{z}^\prime)^\top P^{-1}(\bm{z}-\bm{z}^\prime) \right) },\, P=\diag(\varphi_2^2,\ldots,\varphi_{1+n_z}^2)$\\
\bottomrule
\end{tabular} 
\end{center}
\caption{Overview of some commonly used kernel functions.\label{tab:kernel}}
\end{table}
\subsection{Hyperparameter Optimization}
\label{sec2:sec:hypopt}
In addition to the selection of a kernel function, values for any hyperparameter must be determined to perform the regression. The number of hyperparameters depends on the kernel function used. We concatenate all hyperparameters in a vector~$\bm{\varphi}$ with size~$n_\varphi\in\N$, where~$\bm{\varphi}\in\Phi\subseteq\R^{n_\varphi}$. The hyperparameter set~$\Phi$ is introduced to cover the different spaces of the individual hyperparameters as defined in the following.
\begin{defn}
	\label{sec2:defn:hyp}
	The set~$\Phi$ is called a hyperparameter set for a kernel function~$k$ if and only if the set~$\Phi$ is a domain for the hyperparameters~$\bm{\varphi}$ of~$k$.
\end{defn}
Often, the signal noise~$\sigma_n^2$, see~\cref{sec2:for:joint_dist}, is also treated as hyperparameter. For a better understanding, we keep the signal noise separated from the hyperparameters. There exist several techniques that allow computing the hyperparameters and the signal noise with respect to one optimality criterion. From a Bayesian perspective, we want to find the vector of hyperparameters~$\bm{\varphi}$ which are most likely for the output data~$Y$ given the inputs~$X$ and a GP model. For this purpose, one approach is to optimize the \emph{log marginal likelihood function} of the GP. Another idea is to split the training set into two disjoint sets, one which is actually used for training, and the other, the validation set, which is used to monitor performance. This approach is known as \emph{cross-validation}. In the following, these two techniques for the selection of hyperparameters are presented.
\subsubsection{Log Marginal Likelihood Approach}
A very common method for the optimization of the hyperparameters is by means of the \emph{negative log marginal likelihood function}, often simply named as (neg. log) likelihood function. It is \emph{marginal} since it is obtained through marginalization over the function~$f_{\text{GP}}$. The marginal likelihood is the likelihood that the output data~$Y\in\R^{{n_\D}}$ fits to the input data~$X$ with the hyperparameters~$\bm\varphi$. It is given by
\begin{align}
	\log p(Y\vert X,\bm{\varphi})=-\frac{1}{2}Y^\top (K+\sigma_n^2 I_{n_\D})^{-1}Y-\frac{1}{2}\log\vert K+\sigma_n^2 I_{n_\D}\vert-\frac{{n_\D}}{2}\log 2\pi.\label{sec2:for:log}
\end{align}
A detailed derivation can be found in~\cite{rasmussen2006gaussian}. The three terms of the marginal likelihood in~\cref{sec2:for:log} have the following roles: 
\begin{itemize}
	\item~$\frac{1}{2}Y^\top (K+\sigma_n^2 I_{n_\D})^{-1}Y$ is the only term that depends on the output data~$Y$ and represents the data-fit.
	\item~$\frac{1}{2}\log\vert K+\sigma_n^2 I_{n_\D}\vert$ penalizes the complexity depending on the kernel function and the input data~$X$.
	\item~$\frac{{n_\D}}{2}\log 2\pi$ is a normalization constant.
\end{itemize}
\begin{rem}
	For the sake of notational simplicity, we suppress the dependency on the hyperparameters of the kernel function~$k$ whenever possible.
\end{rem}
The optimal hyperparameters~$\bm{\varphi}^*\in\Phi$ and signal noise~$\sigma_n^*$ in the sense of the likelihood are obtained as the minimum of the negative log marginal likelihood function
\begin{align}
	\begin{bmatrix}
	\bm{\varphi}^*\\
	\sigma_n^*
	\end{bmatrix}=\arg\min_{\bm{\varphi}\in\Phi,\sigma_n\in\R_{\geq 0}}\log p(Y\vert X,\bm{\varphi}).\label{sec2:for:logopt}
\end{align}
Since an analytic solution of the derivation of~\cref{sec2:for:log} is impossible, a gradient based optimization algorithm is typically used to minimize the function. However, the negative log likelihood is non-convex in general such that there is no guarantee to find the optimum~$\bm{\varphi}^*,\sigma_n^*$. In fact, every local minimum corresponds to a particular interpretation of the data. In the following example, we visualize how the hyperparameters affect the regression.
\begin{exam}
A GPR with the squared exponential kernel is trained on eight data points. The signal variance is fixed to~$\varphi_{1}=2.13$. First, we visualize the influence of the lengthscale. For this purpose, the signal noise is fixed to~$\sigma_n=0.21$.~\Cref{sec2:fig:model_selection2} shows the posterior mean of the regression and the neg. log likelihood function. On the left side are three posterior means for different lengthscales. A short lengthscale results in overfitting whereas a large lengthscale smooths out the training data (black crosses). The dotted red function represents the mean with optimized lengthscale by a descent gradient algorithm with respect to~\cref{sec2:for:logopt}. The right plot shows the neg. log likelihood over the signal variance~$\varphi_{1}$ and lengthscale~$\varphi_{2}$. The minimum is here at~$\bm{\varphi}^*=[2.13,1.58]^\top$.
	\begin{center}
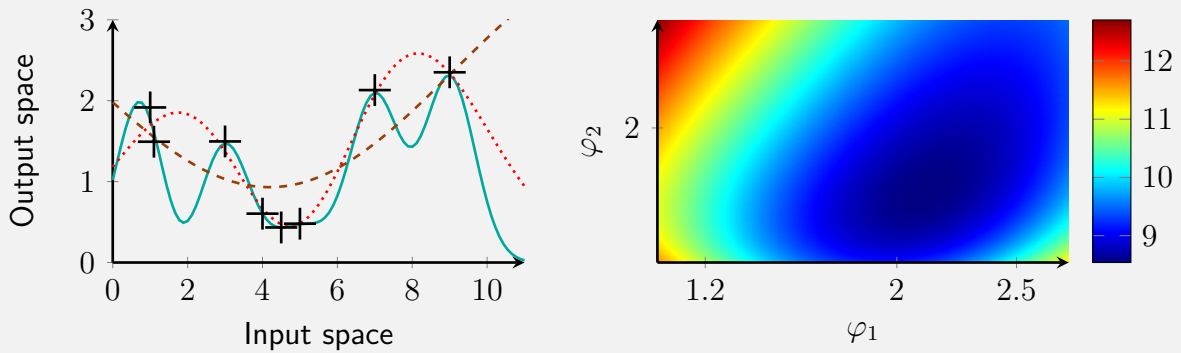

		\tikzsetnextfilename{section2_model_selection2}
		\vspace{-0.2cm}
		\captionsetup{type=figure}\begin{tikzpicture}
	\begin{axis}[
	  name=plot1,
	  axis lines=left,
	  xlabel={Input space},
	  ylabel={Output space},
	  font={\sffamily},
	  line width=1pt,
	  grid = none,
	  height=4.8cm,
	  width=7cm,
	  xmin=0, xmax=11,ymin=0,ymax=3,
	 ylabel style={at={(-0.15,0.5)}}]
			 \addplot[mark=+,color=black, only marks,mark size=6,line width=1pt] table [x index=0,y index=1]{data/section2/model_selection2_example_data.dat};
  			 \addplot[color=Emerald,line width=1pt] table [x index=0,y index=1]{data/section2/model_selection2_example_mean_var.dat};
  			 \addplot[color=RawSienna,dashed,line width=1pt] table [x index=0,y index=3]{data/section2/model_selection2_example_mean_var.dat};	 
  			 \addplot[color=red,dotted,line width=1pt] table [x index=0,y index=5]{data/section2/model_selection2_example_mean_var.dat};	 
	\end{axis}
	\begin{axis}[
	  name=plot2,
	  xshift=-1mm,
	  at=(plot1.right of south east), anchor=left of south west,
	  xshift=0.7cm,
	  axis lines=left,
	  font={\sffamily},
	  xlabel={$\varphi_{1}$},
	  ylabel={$\varphi_{2}$},
	  xtick={1.2,2,2.5},
	  line width=1pt,
	  view={0}{90}, 
      colormap/jet, 
      shader=interp,
	  grid = none,
	  colorbar,
	  height=4.8cm,
	  width=7cm]
			 \addplot3[surf,mesh/cols=40] table {data/section2/model_selection2_hyps.dat};		 
	\end{axis}
	\end{tikzpicture} 
		\vspace{-0.7cm}
		\captionof{figure}{Left: Regression with different lengthscales:~$\varphi_{2}=0.67$ (cyan, solid),~$\varphi_{2}=7.39$ (brown, dashed), and~$\varphi_{2}=1.58$ (red, dotted). Right: Neg. log likelihood function over signal variance~$\varphi_{1}$ and lengthscale~$\varphi_{2}$.}
		\label{sec2:fig:model_selection2}
	\end{center}
Next, the meaning of different interpretations of the data is visualized by varying the signal noise~$\sigma_n$ and the lengthscale~$\varphi_{2}$. The right plot of~\cref{sec2:fig:model_selection3} shows two minima of the negative log likelihood function. The lower left minimum at~$\log(\sigma_n)=0.73$ and~$\log(\varphi_{2})=-1.51$ interprets the data as slightly noisy which leads to the dotted red posterior mean in the left plot. In contrast, the upper right minimum at~$\log(\sigma_n)=5$ and~$\log(\varphi_{2})=-0.24$ interprets the data as very noisy without a trend, which manifests as the cyan posterior mean in the left plot. Depending on the initial value, a gradient based optimizer would terminate in one of these minima.
		\begin{center}
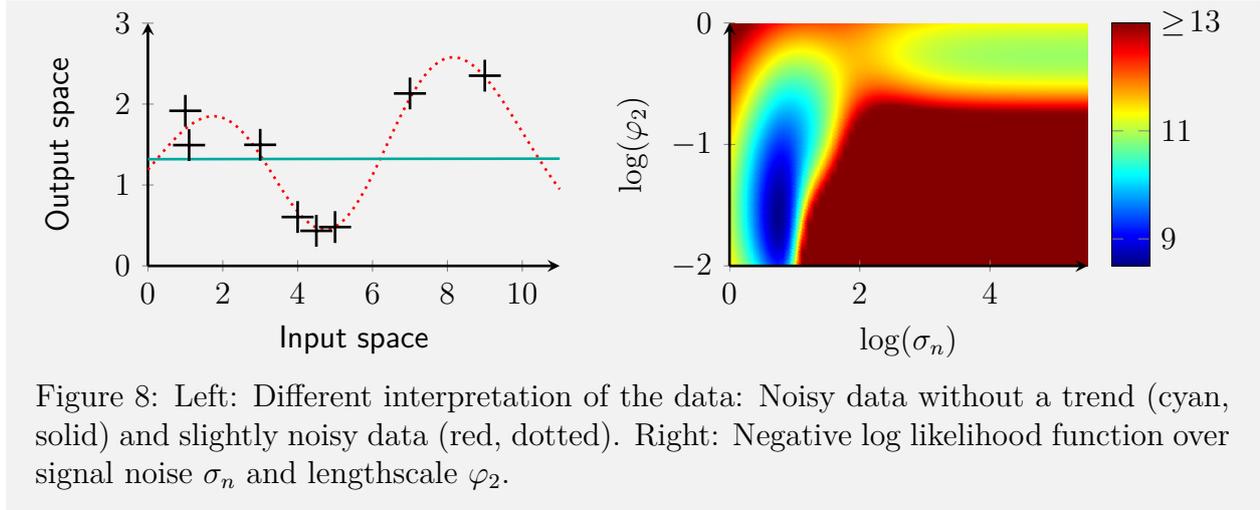

		\tikzsetnextfilename{section2_model_selection3}
		\vspace{-0.4cm}
		\captionsetup{type=figure}\begin{tikzpicture}
	\begin{axis}[
	  name=plot1,
	  axis lines=left,
	  xlabel={Input space},
	  ylabel={Output space},
	  font={\sffamily},
	  line width=1pt,
	  grid = none,
	  height=4.8cm,
	  width=7cm,
	  xmin=0, xmax=11,ymin=0,ymax=3,
	 ylabel style={at={(-0.15,0.5)}}]
			 \addplot[mark=+,color=black, only marks,mark size=6,line width=1pt] table [x index=0,y index=1]{data/section2/model_selection3_example_data.dat};
  			 \addplot[color=red,dotted,line width=1pt] table [x index=0,y index=1]{data/section2/model_selection3_example_mean_var.dat};
  			 \addplot[color=Emerald,line width=1pt] table [x index=0,y index=3]{data/section2/model_selection3_example_mean_var.dat};	  
	\end{axis}
	\begin{axis}[
	  name=plot2,
	  xshift=-1mm,
	  at=(plot1.right of south east), anchor=left of south west,
	  xshift=0.7cm,
	  axis lines=left,
	  font={\sffamily},
	  xlabel={$\log(\sigma_n)$},
	  ylabel={$\log(\varphi_{2})$},
	  line width=1pt,
	  point meta min=8.5,
	  point meta max=13,
	  view={0}{90}, 
      colormap/jet, 
      shader=interp,
	  grid = none,
	  colorbar,
	  colorbar style={ytick={9,11,13},yticklabels={9,11,$\geq\!13$}},
	  height=4.8cm,
	  width=6.3cm]
			 \addplot3[surf,mesh/cols=60] table {data/section2/model_selection3_hyps.dat};		 
	\end{axis}
	\end{tikzpicture} 
		\vspace{-0.7cm}
		\captionof{figure}{Left: Different interpretation of the data: Noisy data without a trend (cyan, solid) and slightly noisy data (red, dotted). Right: Negative log likelihood function over signal noise~$\sigma_n$ and lengthscale~$\varphi_{2}$.}
		\label{sec2:fig:model_selection3}
	\end{center}
\end{exam}
\subsubsection{Cross-validation Approach}
This approach works with a separation of the data set~$\mathcal{D}$ in two classes: one for training and one for validation. Cross-validation is almost always used in the~$k_{\text{cv}}$-fold cross-validation setting: the~$k_{\text{cv}}$-fold cross-validation data is split into~$k_{\text{cv}}$ disjoint, equally sized subsets; validation is done on a single subset and training is done using the union of the remaining~$k_{\text{cv}}-1$ subsets, the entire procedure is repeated~$k_{\text{cv}}$ times, each time with a different subset for validation. Here, without loss of generality, we present the leave-one-out cross-validation, which means~$k_{\text{cv}}={n_\D}$. The predictive log probability when leaving out a training point~$\{\x_\text{dat}^{\{i\}},\tilde{y}_\text{dat}^{\{i\}}\}$ is given by
\begin{align}
	\log p(y_\text{dat}^{\{i\}}\vert X,Y_{-i},\bm{\varphi})=-\frac{1}{2}\log\left(\var_{-i}\right)-\frac{\left(\tilde{y}_\text{dat}^{\{i\}}-\mean_{-i}\right)^2}{2\var_{-i}} -\frac{{n_\D}}{2}\log 2\pi,\label{sec2:for:cross}
\end{align}
where~$\mean_{-i}=\mean(f_\text{GP}(\x_\text{dat}^{\{i\}})\vert \x_\text{dat}^{\{i\}},X_{:,-i},Y_{-i})$ and~$\var_{-i}=\var(f_\text{GP}(\x_\text{dat}^{\{i\}})\vert \x_\text{dat}^{\{i\}},X_{:,-i},Y_{-i})$. The~$-i$ index indicates~$X$ and~$Y$ without the element~$\x_\text{dat}^{\{i\}}$ and~$\tilde{y}_\text{dat}^{\{i\}}$, respectively. Thus,~\cref{sec2:for:cross} is the probability for the output~$y_\text{dat}^{\{i\}}$ at~$\x_\text{dat}^{\{i\}}$ but without the training point~$\{\x_\text{dat}^{\{i\}},\tilde{y}^{\{i\}}\}$. Accordingly, the leave-one-out log predictive probability~$L_{\text{LOO}}\in\R$ is
\begin{align}
	L_{\text{LOO}}=\sum_{i=1}^{n_\D} \log p(y_\text{dat}^{\{i\}}\vert X,Y_{-i},\bm{\varphi}).
\end{align}
In comparison to the log likelihood approach~\cref{sec2:for:logopt}, the cross-validation is in general more computationally expensive but might find a better representation of the data set, see~\cite{geisser1979predictive} for discussion and related approaches.
\section{Gaussian Process Dynamical Models}\label{sec2:sec:GPDM}
So far, we consider GPR in non-dynamical settings where only an input-to-output mapping is considered. However, Gaussian process dynamical models (GPDMs) have recently become a versatile tool in system identification because of their beneficial properties such as the bias variance trade-off and the strong connection to Bayesian mathematics, see \cite{frigola2014variational}. In many works, where GPs are applied to dynamical model, only the mean function of the process is employed, e.g., in \cite{wang2005gaussian} and \cite{chowdhary2013bayesian}. This is mainly because GP models are often used to replace deterministic parametric models. However, GPDMs contain a much richer description of the underlying dynamics, but also the uncertainty about the model itself when the full probabilistic representation is considered. Therefore, one main aspect of GPDMs is to distinguish between recurrent structures and non-recurrent structures. A model is called recurrent if parts of the regression vector depend on the outputs of the model. Even though recurrent models become more complex in terms of their behavior, they allow to model sequences of data, see~\cite{sjoberg1995nonlinear}. If all states are fed back from the model itself, we get a simulation model, which is a special case of the recurrent structure. The advantage of such a model is its property to be independent from the real system. Thus, it is suitable for simulations, as it allows multi-step ahead predictions. In this report, we focus on two often-used recurrent structures: the Gaussian process state space model (GP-SSM) and the Gaussian process nonlinear error output (GP-NOE) model.
\subsection{Gaussian Process State Space Models}
\begin{figure}[h]
	\begin{center}
		\tikzsetnextfilename{section2_model_structure_ssm}
		\begin{tikzpicture}[node distance=2.5cm,auto,>=latex]
	\def\nodedist{2.5cm}
	\def\muxerheight{3cm}
	\tikzstyle{mux}=[draw, fill=black, minimum height=\muxerheight,minimum width=0.05cm,inner sep=0pt]
	\tikzstyle{block}=[draw, minimum height=0.5cm,minimum width=0.5cm,inner sep=2pt]
	\tikzstyle{textblock}=[text width=0.5cm, anchor=center,align=center,inner sep=0pt]
    \node [mux] (muxer1) {};
    \path (muxer1.west)+(-\nodedist*1.5,-\muxerheight/4) node (input_u) [coordinate] {};
    \path (muxer1.west)+(-\nodedist/1.5,\muxerheight/4) node (q) [block] {$\mathfrak{P}^{-1}$};
    \path (q.west)+(-\nodedist/8,0) node (qc) [coordinate] {};
    \path (muxer1.east)+(\nodedist/10,0) node (cord1) [coordinate] {};
    \path (muxer1.east)+(\nodedist/2,\muxerheight/8*3) node (GP1) [block] {$\GP$};
    \path (muxer1.east)+(\nodedist/2,\muxerheight/8*1) node (GP2) [block] {$\GP$};
    \path (muxer1.east)+(\nodedist/2,-\muxerheight/8*1) node (text1) [textblock] {\rvdots};
    \path (muxer1.east)+(\nodedist/2,-\muxerheight/8*3) node (GPn) [block] {$\GP$};
    \path (muxer1.east)+(\nodedist/1,0) node (muxer2) [mux] {};
    \path (muxer2.east)+(\nodedist/1.5,0) node (output_x) [coordinate] {};
    \path (output_x)+(\nodedist*0.8,0) node (GPy1) [block] {$p(\ykp\vert\xkp,\bm{\gamma}_y)$};
    \path (GPy1.east)+(\nodedist*0.7,0) node (output_y) [coordinate] {};
    \path[->] (q) edge node [pos=.4,yshift=-1mm] {$\xk$} (muxer1.west |- q.east);
    \path[->] (input_u) edge node [pos=.1,yshift=-1mm] {$\uk$} (muxer1.west |- input_u);
    \draw[->] (muxer1) -- (cord1) |- (GP1);
    \draw[->] (muxer1) -- (cord1) |- (GP2);
    \draw[->] (muxer1) -- (cord1) |- (GPn);
    %
    %
    \path[->] (GP1) edge (muxer2.west |- GP1);
    \path[->] (GP2) edge (muxer2.west |- GP2);
    \path[->] (GPn) edge (muxer2.west |- GPn);
    \draw[->] (muxer2) -- node [pos=.7,yshift=-1mm] {$\xkp$} (GPy1);
    \draw[->] (GPy1) -- node [pos=.5,yshift=-1mm] {$\ykp$} (output_y);
    \draw[->] (GPy1) -| node[below] (a) {} +(-\nodedist*1.2,\muxerheight/1.5) -| node[below] (b) {} (qc) -- (q); 
    \node[draw,dashed,inner xsep=0.3cm,inner ysep=0.5cm,fit=(muxer1) (muxer2) (q) (a) (qc) (b) (GPy1)] (box1)  {};
\end{tikzpicture}
		\caption{Structure of a GP-SSM with~$\mathfrak{P}$ as backshift operator, such that~$\mathfrak{P}^{-1}\xkp\!=\!\xk$}
		\label{fig:MS_SSM}
	\end{center}
\end{figure}
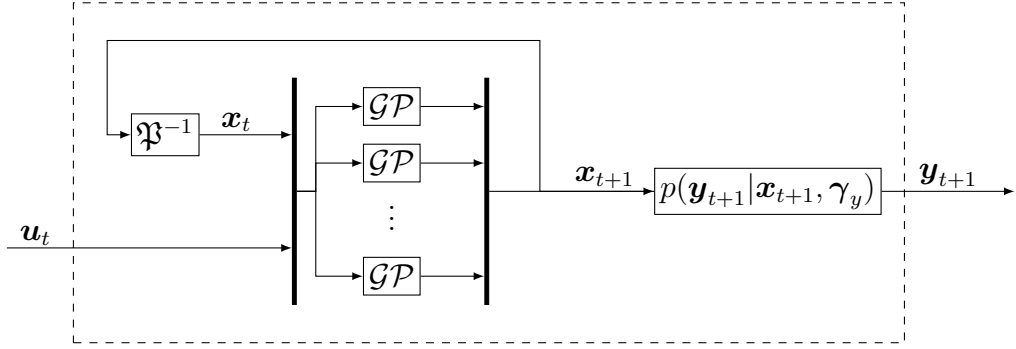

Gaussian process state space models are structured as a discrete-time system. In this case, the states are the regressors, which is visualized in~\cref{fig:MS_SSM}. 
This approach allows to be more efficient, since the regressors are less restricted in their internal structure as in input-output models. Thus, a very efficient model in terms of number of regressors might be possible. The mapping from the states to the output is often be assumed to be known. The situation, where the output mapping describes a known sensor model, is such an example. It is mentioned in~\cite{frigola2013bayesian} that using too flexible models for both, the state mapping~$\dyn$ and the output mapping, can result in problems of non-identifiability. Therefore, we focus on a known output mapping. The mathematical model of the GP-SSM is thus given by
\begin{align}
\begin{split}
	\xkp&=\dyn(\vxi_t)=\begin{cases} 
f_1(\vxi_t)\sim \GP\left(m^1(\vxi_t),k^1(\vxi_t,\vxi_t^\prime)\right)\\
\vdots\hspace{0.9cm}\vdots\hspace{0.5cm}\vdots\\
f_{n_x}(\vxi_t)\sim \GP\left(m^{n_x}(\vxi_t),k^{n_x}(\vxi_t,\vxi_t^\prime)\right).
\end{cases}\\
	\yk&\sim p(\yk\vert\xk,\bm{\gamma}_y),\label{for:gp_ssm}
	\end{split}
\end{align}
where~$\vxi_t\in\R^{n_\xi},n_\xi=n_x+n_u$ is the concatenation of the state vector~$\xk\in\X\subseteq\R^{n_x}$ and the input~$\uk\in\mathcal{U}\subseteq\R^{n_u}$ such that~$\vxi_t=[{\xk};\inputu_t]$. The mean function is given by continuous functions~$m^1,\ldots,m^{n_x}\colon\R^{n_\xi}\to\R$. The output mapping is parametrized by a known vector~$\bm{\gamma}_y\in\R^{n_\gamma}$ with~$n_\gamma\in\N$. The system identification task for the GP-SSM mainly focuses on~$\dyn$ in particular. It can be described as finding the state-transition probability conditioned on the observed training data.
\begin{rem}
	The potentially unknown number of regressors can be determined using established nonlinear identification techniques as presented in~\cite{keviczky1999nonlinear}, or exploiting embedded techniques such as automatic relevance determination~\cite{kocijan2016modelling}. A mismatch leads to similar issues as in parametric system identification.
\end{rem}
\subsection{Gaussian Process Nonlinear Output Error Models}
\begin{figure}[h]
	\begin{center}
		\tikzsetnextfilename{section2_model_structure_noe}
		\begin{tikzpicture}[node distance=3cm,auto,>=latex]
	\def\nodedist{2.5cm}
	\def\muxerheight{3cm}
	\def\muxerheightb{3cm}
	\tikzstyle{mux}=[draw, fill=black, minimum height=\muxerheight,minimum width=0.05cm,inner sep=0pt]
	\tikzstyle{block}=[draw, minimum height=0.5cm,minimum width=0.5cm,inner sep=2pt]
	\tikzstyle{textblock}=[text width=0.5cm, anchor=center,align=center,inner sep=0pt]
    \node [mux,minimum height=\muxerheightb] (muxer1) {};
    \path (muxer1.west)+(-\nodedist*3,-\muxerheightb/10) node (input_u) [coordinate] {};
    \path (muxer1.west)+(-\nodedist,-\muxerheightb/10*2.5) node (text2) [textblock] {\rvdots};
    \path (muxer1.west)+(-\nodedist*3,-\muxerheightb/10*4) node (input_un) [coordinate] {};
    \path (muxer1.west)+(-\nodedist*2,\muxerheightb/10*4) node (q) [block] {$\mathfrak{P}^{-\!1}$};
    \path (muxer1.west)+(-\nodedist,\muxerheightb/10*2.8) node (text3) [textblock] {\rvdots};
    \path (muxer1.west)+(-\nodedist,\muxerheightb/10) node (qn) [block] {$\mathfrak{P}^{-\!n_\text{out}}$};
    \path (q.west)+(-\nodedist/8,0) node (qc) [coordinate] {};
    \path (muxer1.east)+(\nodedist/6,0) node (cord1) [coordinate] {};
    \path (muxer1.east)+(\nodedist*0.6,\muxerheight/8*3) node (GP1) [block] {$\GP$};
    \path (muxer1.east)+(\nodedist*0.6,\muxerheight/8*1) node (GP2) [block] {$\GP$};
    \path (muxer1.east)+(\nodedist*0.6,-\muxerheight/8*1) node (text1) [textblock] {\rvdots};
    \path (muxer1.east)+(\nodedist*0.6,-\muxerheight/8*3) node (GPn) [block] {$\GP$};
    \path (muxer1.east)+(\nodedist*1.2,0) node (muxer2) [mux] {};
    \path (muxer2.east)+(\nodedist,0) node (output_x) [coordinate] {};
    \path[->] (q) edge node [yshift=-1mm,pos=.3,text width=0.6cm,align=left] {$\yk$} (muxer1.west |- q.east);
    \path[->] (qn) edge node [yshift=-1.2mm,pos=.25,text width=0.6cm,align=left] {$\y_{t\!-\!n_\text{out}\!+\!1}$} (muxer1.west |- qn.east);
    \path[->] (input_u) edge node [yshift=-1mm,pos=.05,text width=0.6cm,align=left] {$\uk$} (muxer1.west |- input_u);
    \path[->] (input_un) edge node [yshift=-1mm,pos=.05,text width=0.6cm,align=left] {$\inputu_{t\!-\!n_\text{in}\!+\!1}$} (muxer1.west |- input_un);
    \draw[->] (muxer1) -- (cord1) |- (GP1);
    \draw[->] (muxer1) -- (cord1) |- (GP2);
    \draw[->] (muxer1) -- (cord1) |- (GPn);
    \path[->] (GP1) edge (muxer2.west |- GP1);
    \path[->] (GP2) edge (muxer2.west |- GP2);
    \path[->] (GPn) edge (muxer2.west |- GPn);
    \draw[->] (muxer2) -- node [yshift=-1mm,pos=.8] {$\ykp$} (output_x);
    \draw[->] (output_x) -| node[below] (a) {} +(-\nodedist*0.8,\muxerheightb/1.6) -| node[below] (b) {} (qc) -- (q); 
    \draw[->] (qc) |- (qn);
    \node[draw,dashed,inner xsep=0.05cm,inner ysep=0.2cm,fit=(muxer1) (muxer2) (q) (a) (qc) (b)] {};
\end{tikzpicture}
		\vspace{0cm}\caption{Structure of a GP-NOE model with~$\mathfrak{P}$ as backshift operator ($\mathfrak{P}^{-1}\ykp=\yk$)}\vspace{-0.7cm}
		\label{fig:MS_NOE}
	\end{center}
\end{figure}
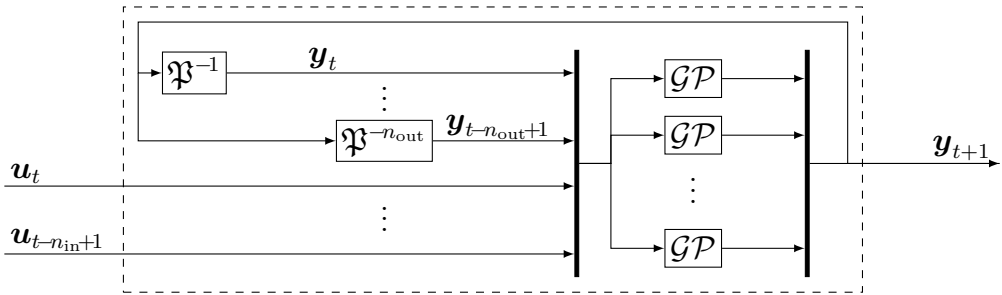

The GP-NOE model uses the past~$n_\text{in}\in\N_{>0}$ input values~$\uk\in\mathcal{U}$ and the past~$n_\text{out}\in\N_{>0}$ output values~$\yk\in\R^{n_y}$ of the model as the regressors.~\Cref{fig:MS_NOE} shows the structure of GP-NOE, where the outputs are feedbacked. Analogously to the GP-SSM, the mathematical model of the GP-NOE is given by
\begin{align}
	&\ykp=\h(\zk)=\begin{cases} 
h_1(\zk)\sim \GP\left(m^1(\zk),k^1(\zk,\zk^\prime)\right)\\
\vdots\hspace{0.9cm}\vdots\hspace{0.5cm}\vdots\\
h_{n_y}(\zk)\sim \GP\left(m^{n_y}(\zk),k^{n_y}(\zk,\zk^\prime)\right),
\end{cases}  \label{for:gp_noe}
\end{align}
where~$\zk\in\R^{n_\zeta},n_\zeta=n_\text{out} n_y+n_\text{in} n_u$ is the concatenation of the past outputs~$\yk$ and inputs~$\uk$ such that~$\zk=[\y_{t-n_\text{out}+1};\ldots;\y_t;\inputu_{t-n_\text{in}+1};\ldots;\inputu_t]$. 
The mean function is given by continuous functions~$m^1,\ldots,m^{n_y}\colon\R^{n_\zeta}\to\R$. In contrast to nonlinear autoregressive exogenous models, that focus on one-step ahead prediction, a NOE model is more suitable for simulations as it considers the multi-step ahead prediction~\cite{nelles2013nonlinear}. However, the drawback is a more complex training procedure that requires a nonlinear optimization scheme due to their recurrent structure~\cite{kocijan2016modelling}.
\FloatBarrier
\begin{rem}
It is always possible to convert an identified input-output model into a state-space model, see\cite{phan1970relationship}. However, focusing on state-space models only would preclude the development of a large number of useful identification results.
\end{rem}
\begin{rem}
Control relevant properties of GP-SSMs and GO-NOE models are discussed in~\cite{beckers2016,beckers:cdc2016,beckers2020}.
\end{rem}
\newpage
\section{Summary}
In this article, we introduce the GP and its usage in GPR. Based on the property, that every finite subset of a GP follows a multi-variate Gaussian distribution, a closed-form equation can be derived to predict the mean and variance for a new test point. The GPR can intrinsically handle noisy output data if it is Gaussian distributed. As GPR is a data-driven method, only little prior knowledge is necessary for the regression. Further, the complexity of GP models scales with the number of training points. A degree of freedom in the modeling is the selection of the kernel function and its hyperparameters. We present an overview of common kernels and the necessary properties to be a valid kernel function. For the hyperparameter determination, two approaches based on numerical optimization are shown. The kernel of the GP is uniquely related to a RKHS, which determines the shape of the samples of the GP. Based on this, we compare different approaches for the quantification of the model error that quantifies the error between the GPR and the actual data-generating function. Finally, we introduce how GP models can be used as dynamical systems in GP-SSMs and GP-NOE models.

	\newpage
	\appendix

\section{Conditional Distribution}\label{app:1}
Let $\bm{\nu}_1\in\R^{n_{\nu_1}},\bm{\nu}_2\in\R^{n_{\nu_2}}$ with $n_{\nu_1},n_{\nu_1}\in\N$ be probability variables, which are multivariate Gaussian distribution
\begin{align}
\begin{bmatrix} \bm{\nu}_1\\\bm{\nu}_2 \end{bmatrix}\sim \mathcal{N} \left(\begin{bmatrix}\Mean_1\\\Mean_2\end{bmatrix}, \begin{bmatrix}\Sigma_{11}&\Sigma_{12}^\top\\ \Sigma_{12}&\Sigma_{22} \end{bmatrix}\right)
\end{align}
with mean~$\Mean_1\!\in\R^{n_{\nu_1}},\Mean_2\!\in\R^{n_{\nu_2}}$ and variance~$\Sigma_{11}\!\in\R^{n_{\nu_1} \times n_{\nu_1}},\Sigma_{12}\!\in\R^{n_{\nu_2} \times n_{\nu_1}},\Sigma_{22}\!\in\R^{n_{\nu_2} \times n_{\nu_2}}$. The task is to determine the conditional probability
\begin{align}
\prob(\bm{\nu}_2|\bm{\nu}_1)&=\frac{\prob(\bm{\nu}_1,\bm{\nu}_2)}{\prob(\bm{\nu}_1)}.
\end{align}
The joined probability $\prob(\bm{\nu}_1,\bm{\nu}_2)$ is a multivariate Gaussian distribution with
\begin{align}
\prob(\bm{\nu}_1,\bm{\nu}_2)&=\frac{1}{(2\pi)^{(n_{\nu_1}+n_{\nu_2})/2}\det(\Sigma)^{\frac{1}{2}}}\exp\left(-\frac{1}{2}(\x-\Mean)^\top\Sigma^{-1}(\x-\Mean)\right)\\
\Mean&\coloneqq\begin{bmatrix}
\Mean_1\\\Mean_2
\end{bmatrix},\quad
\Sigma\coloneqq\begin{bmatrix}
\Sigma_{11}&\Sigma_{12}^\top\\ \Sigma_{12}&\Sigma_{22} 
\end{bmatrix},
\end{align}
where $\x=[\x_1;\x_2],\x_1\in\R^{n_{\nu_1}},\x_2\in\R^{n_{\nu_2}}$. The marginal distribution of $\bm{\nu}_1$ is defined by the mean $\Mean_1$ and the variance $\Sigma_{11}$ such that
\begin{align}
\prob(\bm{\nu}_1)=\frac{1}{(2\pi)^{\frac{n_{\nu_1}}{2}}\det(\Sigma_{11})^{\frac{1}{2}}}\exp\left(-\frac{1}{2}(\x_1-\Mean_1)^\top\Sigma_{11}^{-1}(\x_1-\Mean_1)\right).
\end{align}
The division of the joint distribution by the marginal distribution results again in a Gaussian distribution with
\begin{align}
\prob(\bm{\nu}_2|\bm{\nu}_1)&=\underbrace{\frac{\det(\Sigma_{11})^{\frac{1}{2}}}{(2\pi)^{\frac{n_{\nu_2}}{2}}\det(\Sigma)^{\frac{1}{2}}}}_{*}\exp\Bigl(-\frac{1}{2}\underbrace{\vphantom{\frac{\det(\Sigma_{22})^{\frac{1}{2}}}{(2\pi)^{(n/2)}}}\left[(\x-\Mean)^\top\Sigma^{-1}(\x-\Mean)-(\x_1-\Mean_1)^\top\Sigma_{11}^{-1}(\x_1-\Mean_1)\right]}_{**}\Bigr),\label{app:eq1}
\end{align}
where the first part $*$ can be rewritten as
\begin{align}
*=\frac{1}{(2\pi)^{\frac{n_{\nu_2}}{2}}}\left(\frac{\det(\Sigma_{11})}{\det(\Sigma_{11})\det(\Sigma_{22}-\Sigma_{12}\Sigma_{11}^{-1}\Sigma_{12}^\top)}\right)^{\frac{1}{2}}=\frac{1}{(2\pi)^{\frac{n_{\nu_2}}{2}}\det(\Sigma_{22}-\Sigma_{12}\Sigma_{11}^{-1}\Sigma_{12}^\top)^{\frac{1}{2}}}.
\end{align}
Thus, the covariance matrix $\Sigma_{22|1}$ of the conditional distribution $\prob(\bm{\nu}_2\vert \bm{\nu}_1)$ is given by
\begin{align}
\Sigma_{22|1}=\Sigma_{22}-\Sigma_{12}\Sigma_{11}^{-1}\Sigma_{12}^\top.
\end{align}
For the simplification of the second part $**$ of~\cref{app:eq1}, we exploit the special block structure of $\Sigma$, such that its inverse is given by
\begin{align}
\Sigma&=\begin{bmatrix}\Sigma_{11} & \Sigma_{12}\\ \Sigma_{21}&\Sigma_{22} \end{bmatrix},\qquad\qquad \Sigma^{-1}=\begin{bmatrix}\Sigma_{11}^\prime&\Sigma_{12}^\prime\\ \Sigma_{21}^\prime&\Sigma_{22}^\prime\end{bmatrix}\\
\begin{split}
\Sigma_{11}^\prime&=\Sigma_{11}^{-1}+\Sigma_{11}^{-1} \Sigma_{12} N \Sigma_{21} \Sigma_{11}^{-1}\\
\Sigma_{12}^\prime&=-\Sigma_{11}^{-1} \Sigma_{12} N\\
\Sigma_{21}^\prime&=- N \Sigma_{21} \Sigma_{11}^{-1}\\
\Sigma_{22}^\prime&= N
\end{split}
\end{align}
with $N=( \Sigma_{22}- \Sigma_{21} \Sigma_{11}^{-1} \Sigma_{12})^{-1}$. Thus, we compute $**$ as
\begin{align}
**&=\begin{bmatrix}\x_1-\Mean_1\\ \x_2-\Mean_2\end{bmatrix}^\top\begin{bmatrix}\Sigma_{11}&\Sigma_{12}^\top\\ \Sigma_{12}&\Sigma_{22}\end{bmatrix}^{-1}\begin{bmatrix}\x_1-\Mean_1\\ \x_2-\Mean_2\end{bmatrix}-(\x_1-\Mean_1)^\top\Sigma_{11}^{-1}(\x_1-\Mean_1)\\
&=(\x_2-\Mean_2)^\top\Sigma_{22|1}^{-1}(\x_2-\Mean_2)+2(\x_2-\Mean_2)^\top\left(-\Sigma_{11}^{-1}\Sigma_{12}^\top\Sigma_{22|1}^{-1}\right)(\x_1-\Mean_1)\notag\\
&+(\x_1-\Mean_1)^\top\left(\Sigma_{11}^{-1}+\Sigma_{11}^{-1}\Sigma_{12}^\top\Sigma_{22|1}^{-1}\Sigma_{12}\Sigma_{11}^{-1}\right)(\x_1-\Mean_1)-(\x_1-\Mean_1)^\top\Sigma_{11}^{-1}(\x_1-\Mean_1)\\
&=\big(\x_2-\underbrace{\Mean_2+\Sigma_{12}\Sigma_{11}^{-1}(\x_1-\Mean_1)}_{\Mean_{2|1}}\big)^\top\Sigma_{22|1}^{-1}\big(\x_2-\underbrace{\Mean_2+\Sigma_{12}\Sigma_{11}^{-1}(\x_1-\Mean_1)}_{\Mean_{2|1}}\big)
\end{align}
Finally, the conditional probability is given with the conditional mean $\Mean_{2|1}\in\R^{n_{\nu_2}}$ and the conditional covariance matrix $\Sigma_{22|1}\in\R^{n_{\nu_2}\times n_{\nu_2}}$ by
\begin{align}
p(\bm{\nu}_2|\bm{\nu}_1)&=\frac{1}{(2\pi)^{\frac{n_{\nu_2}}{2}}\det(\Sigma_{22|1})^{\frac{1}{2}}}\exp\left(-\frac{1}{2}(\x_2-\Mean_{2|1})^\top\Sigma_{22|1}^{-1}(\x_2-\Mean_{2|1})\right)\\
\begin{split}
\Mean_{2|1}&=\Mean_2+\Sigma_{12}\Sigma_{11}^{-1}(\x_1-\Mean_1)\\
\Sigma_{22|1}&=\Sigma_{22}-\Sigma_{12}\Sigma_{11}^{-1}\Sigma_{12}^\top.
\end{split}
\end{align}
	
	\clearpage
	
	\vspace{\stretch{1}}
	\printbibliography

\end{document}